\documentclass[aps,prfluids,print,longbibliography,superscriptaddress]{revtex4-2}
\usepackage{hyperref}
\usepackage[utf8]{inputenc}
\usepackage{amsmath}
\usepackage{amsthm}
\usepackage{amssymb}
\usepackage{bm}
\usepackage{mathtools}
\usepackage[tmargin=1in,bmargin=1in,lmargin=1in,lmargin=1in]{geometry}
\usepackage{graphicx}
\graphicspath{{figures/}}
\usepackage[export]{adjustbox}
\usepackage{tabularx}
\usepackage{enumitem}
\usepackage{siunitx}
\usepackage[dvipsnames]{xcolor}
\usepackage{stmaryrd}

\usepackage{bigints}
\usepackage{latexsym}
\usepackage{multirow}
\usepackage{rotating}
\usepackage{caption}
\usepackage{subcaption}
\usepackage{url}

\newcommand{\beq}{\begin{equation}}
\newcommand{\eeq}{\end{equation}}

\def\mathbi#1{\textbf{\em #1}}

\def\Xint#1{\mathchoice
   {\XXint\displaystyle\textstyle{#1}}%
   {\XXint\textstyle\scriptstyle{#1}}%
   {\XXint\scriptstyle\scriptscriptstyle{#1}}%
   {\XXint\scriptscriptstyle\scriptscriptstyle{#1}}%
   \!\int}
\def\XXint#1#2#3{{\setbox0=\hbox{$#1{#2#3}{\int}$}
     \vcenter{\hbox{$#2#3$}}\kern-.5\wd0}}

\def\dashint{\Xint-}

\providecommand\bnabla{\boldsymbol{\nabla}}
\providecommand\bcdot{\boldsymbol{\cdot}}
\DeclareMathOperator{\Tr}{Tr}

\usepackage{xcolor} 
\definecolor{newcolor}{rgb}{.8,.349,.1}

\definecolor{lightblue}{rgb}{0.63, 0.74, 0.78}
\definecolor{seagreen}{rgb}{0.18, 0.42, 0.41}
\definecolor{orange}{rgb}{0.85, 0.55, 0.13}
\definecolor{silver}{rgb}{0.69, 0.67, 0.66}
\definecolor{rust}{rgb}{0.72, 0.26, 0.06}
\definecolor{joshua}{RGB}{251,220,127}

\colorlet{lightsilver}{silver!30!white}
\colorlet{darkorange}{orange!85!black}
\colorlet{darksilver}{silver!85!black}
\colorlet{darklightblue}{lightblue!85!black}
\colorlet{darkrust}{rust!85!black}

\hypersetup{
    colorlinks=true,
    pdffitwindow=true,
    pdfpagelayout=SinglePage
}

\begin{document}

\hypersetup{
    linkcolor=darkrust,       		
    citecolor=seagreen,       		
    filecolor=black,      		
    urlcolor=seagreen,       		
    }

\title{A spectral boundary integral method for simulating electrohydrodynamic flows in viscous drops}


\author{Mohammadhossein Firouznia} 
\affiliation{Department of Mechanical and Aerospace Engineering, University of California San Diego, 9500 Gilman Drive, La Jolla, CA 92093, USA}

\author{Spencer H. Bryngelson} 
\affiliation{School of Computational Science \& Engineering, Georgia Institute of Technology, 756 West Peachtree Street NW, Atlanta, GA 30332, USA}

\author{David Saintillan}
\email[]{Corresponding author: dstn@ucsd.edu}
\affiliation{Department of Mechanical and Aerospace Engineering, University of California San Diego, 9500 Gilman Drive, La Jolla, CA 92093, USA}


\date{\today}
\begin{abstract}
	A weakly conducting liquid droplet immersed in another leaky dielectric liquid can exhibit rich dynamical behaviors under the effect of an applied electric field.
	Depending on material properties and field strength, the nonlinear coupling of interfacial charge transport and fluid flow can trigger electrohydrodynamic instabilities that lead to shape deformations and complex dynamics.
	We present a spectral boundary integral method to simulate droplet electrohydrodynamics in a uniform electric field.
	All physical variables, such as drop shape and interfacial charge density, are represented using spherical harmonic expansions. In addition to its exponential accuracy, the spectral representation affords a nondissipative dealiasing method required for numerical stability. 
	A comprehensive charge transport model, valid under a wide range of electric field strengths, accounts for charge relaxation, Ohmic conduction, and surface charge convection by the flow.
	A shape reparametrization technique enables the exploration of significant droplet deformation regimes.
	For low-viscosity drops, the convection by the flow drives steep interfacial charge gradients near the drop equator.
	This introduces numerical ringing artifacts we treat via a weighted spherical harmonic expansion, resulting in solution convergence.
	The method and simulations are validated against experimental data and analytical predictions in the axisymmetric Taylor and Quincke electrorotation regimes. 
\end{abstract}


\maketitle

\section{Introduction \label{sec:intro}}

A wide range of engineering applications involve liquid drops immersed in another fluid while subject to an applied electric field.
Some examples include ink-jet printing \citep{Basaran2013Ann_rev_inkjet}, electrospraying \citep{fernandez2007Ann_rev_spray}, and microfluidic devices and pumps \citep{laser2004review_micropumps}.
These systems exhibit rich dynamics due to the electric field and fluid flow coupling.
When an interface separating two immiscible fluids is subject to an otherwise uniform electric field, the electric field undergoes a jump across the interface due to the mismatch in material properties.
This discontinuity in the electric field induces electric stresses that can deform the interface and drive the fluid into motion.

We are interested in leaky dielectric liquids such as oils, which serve as poor conductors.
Unlike electrolyte solutions where diffuse Debye layers affect the system's dynamics, leaky dielectrics are characterized by the absence of diffuse Debye layers \citep{saville1997LDM_Ann_Rev}.
The free charges instead concentrate on the interfaces between different phases in the system.
Consequently, the electric field acting on the interfacial charge creates electric stresses along the normal and tangential directions, which cause deformations and fluid motion.
Surface tension has a stabilizing effect in general, trying to restore the equilibrium shape.
Melcher and Taylor developed a framework for studying electrohydrodynamic phenomena in leaky dielectric  systems, known as the leaky dielectric model (LDM) \citep{melcher&taylor1969LDM}.
Central to their work is a charge conservation model that describes a balance between Ohmic fluxes from the bulk, interfacial charge convection, and finite charge relaxation.
It was previously shown that LDM can be derived asymptotically from electrokinetic
models in the limit of strong electric fields and thin Debye layers \citep{schnitzer2015taylor, mori2018electrodiffusion}.

This work focuses on the dynamics of a leaky dielectric drop immersed in another dielectric fluid under a uniform DC electric field.
This canonical problem has been a long-standing research problem in electrohydrodynamics.
In his pioneering work, Taylor \citep{taylor1966LDM} formulated a small-deformation theory for an isolated drop based on LDM and could predict oblate and prolate steady shapes depending on the material properties.
While Taylor's theory shows good agreement with experimental data in the limit of vanishing electric capillary number $\mathrm{Ca}_\mathrm{E}$ (ratio of electric to capillary forces), the discrepancy is significant at larger values of $\mathrm{Ca}_\mathrm{E}$.
Therefore, other researchers attempted to extend Taylor's work by accounting for second-order effects in $\mathrm{Ca}_\mathrm{E}$ \citep{ajayi1978note},  considering spheroidal drops \citep{zabarankin2013liquid, zhang2013transient}, including inertial effects \citep{lanauze2013inertia_EHD_drop} and interfacial charge convection \citep{shkadov2002drop, feng2002_2d_EHD, he_salipate2013electrorotation, das2017nonlinear}.

{A variety of computational models have been developed to study drop dynamics under strong electric fields at finite deformations, a problem untractable using analytical theories.
	In the limit of negligible inertia, boundary integral equations can be used to formulate and solve the coupled electrohydrodynamic problem.
	Sherwood was the first to develop a boundary element method for an axisymmetric drop in an equiviscous system and applied it to capture breakup modes in prolate drops \citep{sherwood1988breakup}.
	His original work was subsequently extended to study drop pair interaction \citep{baygents1998EHD} and to cover a wider range of fluid and electric parameters \citep{lac2007axisymmetric}. These earlier attempts used a simplified boundary condition for the electric problem, which neglected transient charge relaxation and interfacial charge convection by the flow. These two effects have recently been shown to play a significant role in drop dynamics and deformations \cite{das2017nonlinear}. Lanauze et al.~\citep{lanauze2015EHD_JFM} and Das and Saintillan \citep{das2017EHD_simulation} recently addressed this problem and developed axisymmetric and three-dimensional boundary element methods based on the full Melcher-Taylor LDM. The effect of charge convection was specifically addressed in \citep{das2017EHD_simulation}, where it was shown to be responsible for Quincke electrorotation. These methods, however, were found to lack accuracy and stability in the regime of strong electric fields. Other numerical approaches have been used to study drop electrohydrodynamics, including immersed boundary \citep{Hu2015IBM_EHD_drop}, level set \citep{bjorklund2009level,theillard2019}, and finite element methods \citep{feng1996FEM_EHD_drop, feng1999EHD_FEMdrop, supeene2008FEM_EHDdrop}. More recently, finite element simulations \citep{collins2013universal_scalinglaws,wagoner2021EHD_lenticular} were also used to investigate electrohydrodynamic instabilities such as tip and equatorial streaming in drops under strong electric fields. These latter techniques all include finite fluid inertia and, with few exceptions \cite{theillard2019}, do not treat the drop surface as a sharp interface. }

{Improved accuracy within the boundary integral framework can be achieved using spectral methods, which rely on expansions of the shape and interfacial variables based on spherical harmonics. Such methods were recently developed to simulate electrohydrodynamics of lipid vesicles \cite{veerapaneni2016BIM_EHD_vesicle} and also extended to the case of individual drops and drop pairs \cite{sorgentone20193d_EHD, sorgentone2021drop_pair, sorgentone2022tandem}. These studies, however, all neglected charge relaxation and charge convection and were thus restricted to weak electric fields. Accurately capturing charge convection is especially challenging as it nonlinearly couples fluid flow and charge transport on a deformed interface. It can result in spurious aliasing errors with negative consequences for accuracy and stability. This work addresses this challenge and presents a spectral boundary integral method for the electrohydrodynamics of deformable liquid drops based on the complete Melcher--Taylor LDM. Interfacial charge convection is rigorously accounted for, and dealiasing and reparametrization techniques are implemented to improve accuracy and stability and enable long-time simulations. }

{The paper is organized as follows.}\ We define the problem and discuss the governing equations and boundary conditions in Sec.~\ref{sec:gov_eqs}, along with their non-dimensionalization in Sec.~\ref{sec:non_dim}.
Sec.~\ref{sec:boundary_int_eqs} presents the integral form of the governing equations and boundary conditions used in developing the boundary integral method.
We discuss different aspects of the numerical method in Sec.~\ref{sec:num_method}:
the spectral representation of all variables in terms of spherical harmonics is discussed in \ref{sec:surf_repres}, followed by details of the dealiasing method in Sec.~\ref{sec:dealias}.
Next, in Sec.~\ref{sec:integration}, we summarize the numerical integration methods used in this study and correction methods to ensure charge neutrality and incompressibility in Sec.~\ref{sec:incompress_charge_neut}.
As explained in Sec.~\ref{sec:reparam}, we also use a reparametrization method to improve the numerical stability in simulations where the drop undergoes significant deformations.
We test and validate our computational model by applying it to a wide range of dynamical behaviors, such as the axisymmetric Taylor regime under weak electric fields in Sec.~\ref{sec:results:taylor_regime}, and Quincke electrorotation under stronger electric fields in Sec.~\ref{sec:results:quincke}.
We also investigate the dynamics of low-viscosity drops in Sec.~\ref{sec:results:low_viscosity_drops}, where charge convection plays an important role.
Finally, we discuss our conclusions and possible extensions of our work in Sec.~\ref{sec:conclusions}. 

\section{Problem definition\label{sec:pb_def}}

\subsection{Governing equations\label{sec:gov_eqs}}

We consider a neutrally buoyant drop of a fluid occupying volume $V^-$ immersed in an infinite body of another fluid $V^+$ while subject to a uniform electric field $\bm{E}_{\infty}=E_{\infty} \,\hat{\bm{e}}_z$ as depicted schematically in Fig.~\ref{figs:schematic}.
The interface $D$ separates the two fluid media, and the surface unit normal $\bm{n}(\bm{x})$ is pointed towards the suspending fluid.
Initially, the drop is uncharged and spherical with radius $r_{0}$.
The material properties, namely the dielectric permittivities, electric conductivities, and dynamic viscosities, are denoted by $(\epsilon^{\pm}, \, \sigma^{\pm}, \, \mu^{\pm})$ inside and outside the drop, respectively.
Under the Taylor--Melcher leaky dielectric model \citep{taylor1966LDM}, any net charge in the system appears on the interface $D$, and the bulk of the fluids remain electroneutral.
Therefore, the electric potential is harmonic in the bulk:
\begin{figure}	
	\centering
	{\includegraphics[width=0.65\textwidth, angle=0]{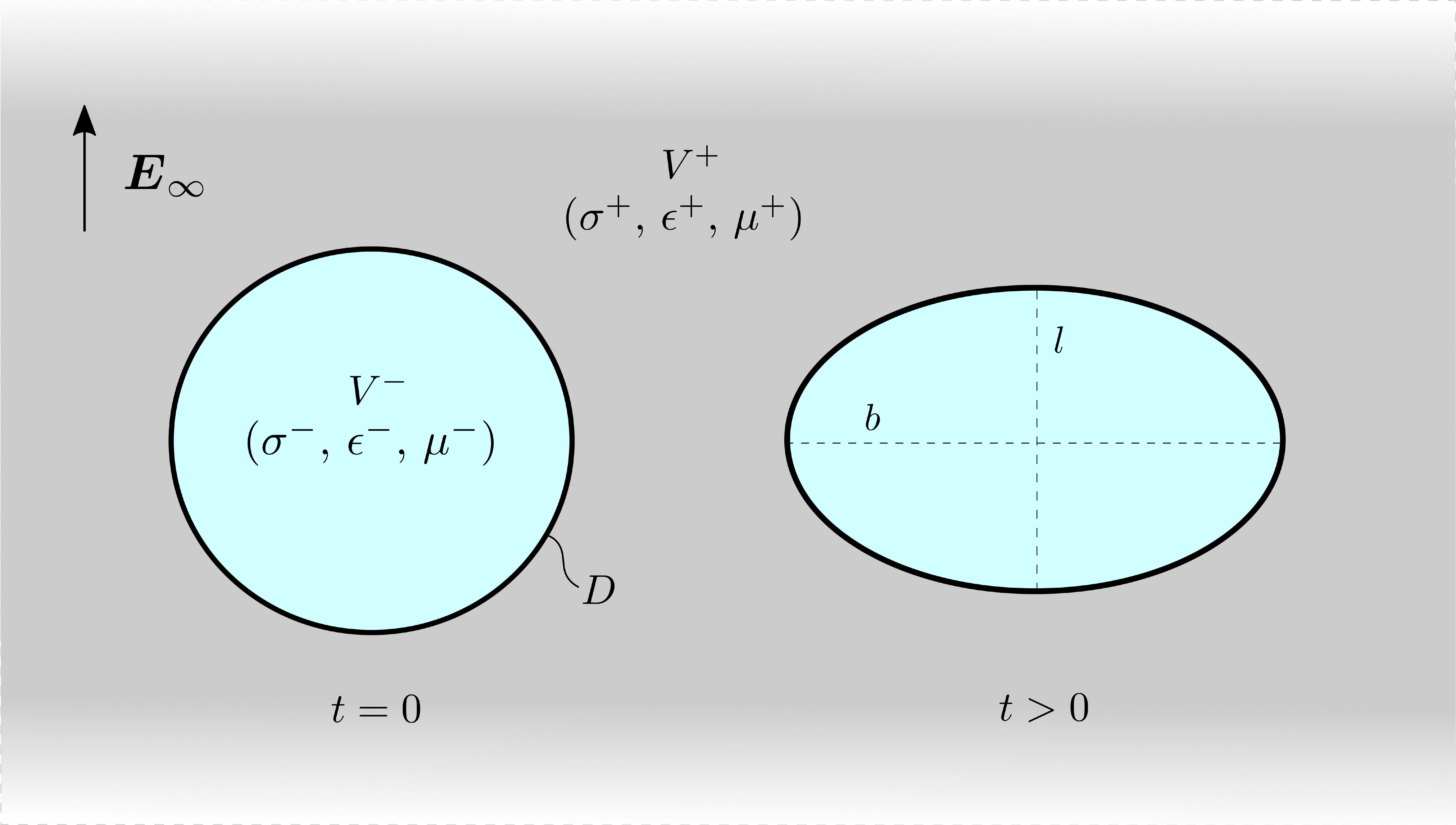} }	
	\caption{
		Problem definition: a leaky dielectric drop with $(\sigma^-,\,\epsilon^-,\, \mu^-)$ is suspended in another leaky dielectric fluid with $(\sigma^+,\,\epsilon^+,\, \mu^+)$ and subject to an external electric field $\bm{E}_\infty$.
		The drop deforms and diverges from its initially spherical shape.
	}  
	\label{figs:schematic}  
\end{figure}

\begin{equation} \label{eq:intro:pot_laplace}
	\nabla^2 \varphi^{\pm}(\bm{x})=0, \qquad   \bm{x}\in V^{\pm}. 
\end{equation} 
Far away from the interface, the electric field $\bm{E}=-\bnabla \varphi $ tends to the applied electric field:
\begin{equation}
	\bm{E}^+\rightarrow \bm{E}_{\infty}=E_{\infty} \, \hat{\bm{e}}_z, \qquad \text{as} ~~ |\bm{x}|\rightarrow \pm \infty. 
\end{equation} 
While the tangential component of the electric field is continuous across the interface, its normal component undergoes a jump due to the mismatch in material properties:
\begin{equation}\label{eq:pb_def:ncrossE_Gausslaw}
	\bm{n}\times\llbracket \bm{E} \rrbracket=\bm{0}, \qquad  \bm{x}\in D.
\end{equation}
We define the operator $\llbracket \mathcal{F} \rrbracket \coloneqq \mathcal{F}^+- \mathcal{F}^-$ as the jump in any variable $\mathcal{F}$ across the interface $D$. 
A surface charge density develops at the interface and follows Gauss's law,
\begin{equation}\label{eq:pb_def:Gausslaw_nondim}
	q(\bm{x})=\bm{n} \bcdot \llbracket \epsilon \bm{E} \rrbracket, \qquad  \bm{x}\in D.
\end{equation}
The surface charge evolves due to Ohmic currents from the bulk and convective currents on the interface. Consequently, it satisfies the conservation equation:
\begin{equation} 
	\partial_t q+\bm{n}\bcdot\llbracket \sigma \bm{E}\rrbracket + \bnabla_s \bcdot(q \bm{u})=0, \qquad \bm{x}\in D, \label{eq:pb_def:chgcons}
\end{equation}
where $\bnabla_s= ({\bm{I}} - \bm{n} \bm{n})\bcdot \bnabla $ is the surface gradient operator and $\bm{u}$ is the fluid velocity.

Neglecting the effect of inertia and gravity, the velocity and pressure fields satisfy the Stokes and continuity equations:
\begin{equation}
	{\mu^{\pm}}\nabla^2\bm{u}^{\pm}-\bnabla p^{\pm}=\mathbf{0},    \qquad   \bnabla \bcdot\bm{u}^{\pm}=0,   \qquad  ~\bm{x}\in V^{\pm}.
\end{equation}
The velocity vector is continuous across the interface and vanishes far from it as
\begin{eqnarray}
	&	\llbracket \bm{u}(\bm{x}) \rrbracket=\bm{0}, &\qquad  \bm{x}\in D, \label{eq:pb_def:cont_vel} \\
	&\bm{u}^+(\bm{x}) \rightarrow \bm{0}, &\qquad \text{as}~~  |\bm{x}| \rightarrow  \infty.\label{eq:pb_def:vel_infty}
\end{eqnarray} 
The balance of interfacial forces requires that the jump in hydrodynamic and electric tractions across the interface balance capillary forces: 
\begin{equation}
	\llbracket \bm{f}^\mathrm{H}\rrbracket + 
	\llbracket \bm{f}^\mathrm{E}\rrbracket=\gamma (\bnabla_s\bcdot \bm{n})\bm{n},
	\qquad \bm{x}\in D. \label{eq:pb_def:dyn_BC}
\end{equation}
We neglect Marangoni effects due to variations in surface tension, $\bnabla _s \gamma = \bm{0}$. 
Hydrodynamic and electric tractions are expressed in terms of the Newtonian and Maxwell stress tensors, respectively: 
\begin{align}
	\bm{f}^\mathrm{H}=\bm{n}\bcdot\bm{T}^\mathrm{H},&\qquad  \bm{T}^\mathrm{H}=-p\bm{I}+\mu\big(\bnabla\bm{u}+{\bnabla\bm{u}}^T\big),\\
	\bm{f}^\mathrm{E}=\bm{n}\bcdot\bm{T}^\mathrm{E},& \qquad \bm{T}^\mathrm{E}=\epsilon \left(\bm{E}\bm{E} -\tfrac{1}{2}E^2\bm{I}\right). \label{eq:pb_def:jump_fH}
\end{align}
The jump in electric tractions can be decomposed into tangential and normal components as
\begin{equation}
	\bm{f}^\mathrm{E} =  \llbracket \epsilon E^{n} \rrbracket \bm{E}^{t}+\frac{1}{2} \llbracket \epsilon ({E^{n}}^2-{E^{t}}^2) \rrbracket \bm{n}=q \bm{E}^{t}+\llbracket p^\mathrm{E} \rrbracket \bm{n}, \label{eq:pb_def:jump_fE_comps}
\end{equation}
where $p^\mathrm{E}=\epsilon({E^{n}}^2 -{E^{t}}^2)/2 $ is the electric pressure \citep{lac2007axisymmetric}. 
The first term on the right-hand side represents the tangential electric stresses in leaky dielectrics, and it vanishes when both fluids are either perfect dielectrics or perfect conductors. 

\subsection{Non-dimensionalization\label{sec:non_dim}}

For the system described above, a dimensional analysis yields five dimensionless groups, three of which characterize the mismatch of material properties in the drop and the suspending fluid:
\beq
\mathrm{R}=\frac{\sigma^+}{\sigma^-}, \qquad \mathrm{Q}=\frac{\epsilon^-}{\epsilon^+},\qquad \mathrm{\lambda}=\frac{\mu^-}{\mu^+}.
\eeq 
The limits of $\mathrm{\lambda} \rightarrow 0$ and $\infty$ correspond to a bubble and a rigid particle, respectively. 
The remaining dimensionless groups describe the system's dynamics and can be obtained by comparing the characteristic time scales in the problem. 
First, note that the response of each fluid phase to Ohmic conduction is characterized by the charge relaxation time:  
\begin{equation}
	\tau_{c}^{\pm}=\frac{\epsilon^{\pm}}{\sigma^{\pm}}.  \label{eq:nondim:RQ}
\end{equation}
The product $\mathrm{RQ}=\tau^-/\tau^+$ is the ratio of charge relaxation times in two fluids and plays an important role in the dynamics of the drop \cite{das2017EHD_simulation}. 
The polarization time for a rigid sphere under an applied electric field is the Maxwell--Wagner relaxation time
\begin{equation}
	\tau_{\mathrm{MW}}=\frac{\epsilon^-+2\epsilon^+}{\sigma^-+ 2\sigma^+}=  \mathrm{\frac{R(Q+2)}{1+2R}} \, \tau_{c}^+, \label{eq:nondim:tau_MW}
\end{equation} 
which provides an approximate timescale for polarization of the drop. 
The accumulation of free charges on the interface creates electric forces that drive the fluid into motion on the electrohydrodynamic time scale 
\begin{equation}
	\tau_{\mathrm{EHD}}=\frac{\mu^+}{\epsilon^+ {E^2_{\infty}}}. \label{eq:nondim:tau_EHD}
\end{equation}
Deformations away from the equilibrium spherical shape relax under the effect of surface tension on the capillary time scale
\begin{equation}
	\tau_{\mathrm{\gamma}}=\frac{\mu^+ \, r_{0}}{\gamma}. \label{eq:nondim:tau_gamma}	
\end{equation}   
By taking the ratios of these time scales, the two remaining dimensionless groups can be defined as
\begin{equation}
	\mathrm{Ca}_{\mathrm{E}}=\frac{\tau_{\gamma}}{\tau_{\mathrm{EHD}}}=\frac{\epsilon {E^2_{\infty}} \, r_{0}}{\gamma}. \quad \mathrm{Ma} =\frac{\tau_{\mathrm{EHD}}}{\tau_{\mathrm{MW}}}=\frac{\mu^+ }{{\tau_{\mathrm{MW}} {\epsilon^+}  E^2_{\infty}} }. \label{eq:nondim:Ca_Ma}
\end{equation}   
The electric capillary number $\mathrm{Ca}_\mathrm{E}$ compares electric forces versus capillary forces, while the Mason number $\mathrm{Ma}$ characterizes the importance of charge conduction against surface charge convection. 
Alternatively, we can construct $\mathrm{Ca}_\mathrm{MW}$ to be independent of the electric field
\begin{equation}
	\mathrm{Ca}_{\mathrm{MW}}=\frac{\tau_{\mathrm{\gamma}}}{\tau_{\mathrm{MW}}}=\frac{\mu^+ (1+\lambda)  \, r_{0}}{\gamma \tau_{\mathrm{MW}}}= (1+\lambda)\, \mathrm{Ca}_\mathrm{E}\,\mathrm{Ma}. \label{eq:nondim:Ca_MW}
\end{equation}  
For a given set of material properties, changing $\mathrm{Ca}_{\mathrm{MW}}$ corresponds to varying the drop radius $r_0$.

We scale the governing equations and boundary conditions using length scale $r_0$, time scale $\tau_{\mathrm{MW}}$, pressure scale $\epsilon^+E^2_{\infty}$, and the characteristic electric potential $E_{\infty}r_0$.
In the remainder of this manuscript, all governing equations and boundary conditions are dimensionless, and results are presented in terms of the corresponding dimensionless variables.

\subsection{Boundary integral formulation\label{sec:boundary_int_eqs}}

The electric problem is formulated in integral form based on the solution to Laplace's equation as \citep{sherwood1988breakup, baygents1998EHD, lac2007axisymmetric}
\begin{equation}\label{eq:BEM:potential_SLP}
	\varphi( \bm{x}_0 )=-\bm{x}_0\bcdot\bm{E}_{\infty} -\int_{D} \bm{n}\bcdot\llbracket \nabla \varphi( \bm{x} ) \rrbracket \,\mathcal{G} \left( \bm{x}_0;\bm{x} \right) \, \mathrm{d}s(\bm{x}),\qquad \text{for}~ \bm{x}_0\in V^{\pm}, D,
\end{equation}
where the evaluation point $\bm{x}_0$ can be anywhere in space, and $\bm{x}$ denotes the integration point on the interface. 
The free-space Green's function for Laplace's equation, $\mathcal{G} \left( \bm{x}_0;\bm{x} \right)$, captures the electric potential due to a point charge in an unbounded domain as
\begin{equation}\label{eq:BEM:greens_laplace}
	\mathcal{G} \left( \bm{x}_0;\bm{x} \right) = \dfrac{1}{4\pi r}, \quad \text{where}~~ \bm{r} = \bm{x}_0-\bm{x}, ~r=|\bm{r}|. 
\end{equation}
Taking the gradient of Eq.~\eqref{eq:BEM:potential_SLP} with respect to $\bm{x}_0$ and using Gauss's law \eqref{eq:pb_def:Gausslaw_nondim}, we derive an integral equation for the jump in the normal electric field as a function of the surface charge distribution:
\begin{equation}
	\dashint_{D} \llbracket E^{n}( \bm{x} ) \rrbracket [\bm{n}(\bm{x}_0)\bcdot\bnabla_0 \mathcal{G}]\,  \mathrm{d}s(\bm{x})-\mathrm{\dfrac{1+Q}{2(1-Q)}}\llbracket E^{n}( \bm{x}_0 ) \rrbracket =E^{n}_{\infty}( \bm{x}_0 )-\dfrac{q(\bm{x}_0)}{1-\mathrm{Q}},\qquad \text{for}~ \bm{x}_0\in D. \label{eq:BEM:integral_eq_En}
\end{equation}
For a given charge distribution $q(\bm{x})$, Eq.~\eqref{eq:BEM:integral_eq_En} can determine $\llbracket E^{n} ( \bm{x} ) \rrbracket $, from which $E^{{n}{+}}$ and $E^{{n}{-}}$ follow as
\begin{equation}
	E^{{n}{+}}( \bm{x} )=\dfrac{q( \bm{x} )-\mathrm{Q} \llbracket E^{n}( \bm{x} ) \rrbracket}{1-\mathrm{Q}}, \qquad  E^{{n}{-}}( \bm{x} )=\dfrac{q( \bm{x} )-\llbracket E^{n}( \bm{x} ) \rrbracket}{1-\mathrm{Q}}. \label{eq:BEM:q_E_in_out}
\end{equation}  
The tangential electric field $\bm{E}^{t}(\bm{x})=-\bnabla_{s} \varphi(\bm{x})$ can be computed by differentiation of the electric potential \eqref{eq:BEM:potential_SLP} along the tangential direction. 
The interfacial jump in electric tractions $\llbracket \bm{f}^\mathrm{E}\rrbracket$ follows from Eq.~\eqref{eq:pb_def:jump_fE_comps} based on the tangential and normal electric fields calculated at every point on the interface. 
This can be used to determine the jump in hydrodynamic tractions  $\llbracket \bm{f}^\mathrm{H}\rrbracket$ using the dynamic boundary condition \eqref{eq:pb_def:dyn_BC} as
\begin{equation}
	\llbracket \bm{f}^\mathrm{H}\rrbracket =- \llbracket \bm{f}^\mathrm{E}\rrbracket+\mathrm{Ca}_\mathrm{E}^{-1} {(\bnabla_{{s}}\bcdot \bm{n})\bm{n}}, 
	\label{eq:BEM:dyn_BC_nondim}
\end{equation}
which enters the calculation of the velocity field, as explained next. 

The flow problem is also recast into a boundary integral form as \citep{rallison1978numerical, pozrikidis1992BIM_book}
\begin{equation}\label{eq:BEM:integral_eq_Stokes}
	\begin{aligned}
		\bm{u}(\bm{x}_0) = & -\dfrac{1}{4\pi \, \mathrm{ {Ma} \, (1+\lambda) }} \int_{D}   \llbracket \bm{f}^\mathrm{H}(\bm{x}) \rrbracket  \bcdot \bm{G}(\bm{x}_0;\bm{x})\,\mathrm{d}s(\bm{x}) \\
		& + \mathrm{ \dfrac{1-\lambda}{4\pi (1+\lambda)} }\,\, \dashint_{D}\bm{u}(\bm{x})\bcdot\bm{T}(\bm{x}_0;\bm{x}) \bcdot \bm{n}(\bm{x})\, \mathrm{d}s(\bm{x}),  \quad \text{for}~ \bm{x}_0\in D. 
	\end{aligned} 
\end{equation} 
Here, $\bm{G}$ is the free-space Green's function for the Stokeslet or flow due to a unit point force in an unbounded domain, and $\bm{T}$ is the corresponding stress tensor:
\begin{equation}\label{eq:BEM:greens_stokes}
	\bm{G}(\bm{x}_0;\bm{x}) = \dfrac{\bm{I}}{r}+\dfrac{\bm{r} \bm{r}}{r^3}, \quad \bm{T}(\bm{x}_0;\bm{x}) = 6 \dfrac{\bm{r} \bm{r} \bm{r}}{r^5}.
\end{equation}
Note that the integral equations \eqref{eq:BEM:potential_SLP}, \eqref{eq:BEM:integral_eq_En} and \eqref{eq:BEM:integral_eq_Stokes} exhibit singular behaviors of different orders as $\bm{x} $  approaches $\bm{x}_0$.
This is due to the singularity of the Green's function for Laplace's \eqref{eq:BEM:greens_laplace} and Stokes equations \eqref{eq:BEM:greens_stokes}.
Conventional quadrature schemes have poor accuracy in the presence of singular integrands and may not converge by increasing the level of discretization.
Therefore, accurate numerical integration requires special treatment of the singularities, which we will discuss in Sec.~\ref{sec:integration}.

\section{Numerical methods\label{sec:num_method}}

We solve Eqs.~\eqref{eq:BEM:integral_eq_En} and \eqref{eq:BEM:integral_eq_Stokes} by building upon a spectral boundary integral method introduced by \citet{zhao_freund2010JCP}.
This method {was also previously applied to study vesicle dynamics under shear and extensional flows \cite{bryngelson18b,bryngelson19a} as well as flows of confined red blood cells \cite{bryngelson19b,bryngelson18a,freund2013flow}.}
All variables, including the interfacial shape, velocity, and charge, are represented using truncated series of spherical harmonic expansions as discussed in Sec.~\ref{sec:surf_repres}.
Nonlinear operations, geometrical quantities (such as mean curvature), and spatial derivatives are computed accurately using a nondissipative dealiasing method discussed in Sec.~\ref{sec:dealias}.
The boundary integrals are computed using
a quadrature scheme for the surface collocation points, with a special treatment for the singular integrands, as summarized in Sec.~\ref{sec:integration}.
Any changes to the volume and net charge of the drop due to numerical errors are corrected as explained in Sec.~\ref{sec:incompress_charge_neut} to ensure numerical stability over long simulation times.
In addition, we adopt a reparameterization technique in Sec.~\ref{sec:reparam} to minimize the high-frequency component of the interfacial shape, which improves the stability of the numerical method for cases with significant deformations.{{The code associated with the methods and simulations in this article is publicly available at \url{https://github.com/mfirouzn/EHD_Drop_3D}}}.

At $t=0$, the drop is uncharged with a spherical shape. The numerical algorithm used in this study follows that of \citet{das2017EHD_simulation}, \citet{firouznia2021PRF_EHD_film}, and \citet{firouznia2022JFM}. 
We perform the following steps at every time iteration:
\begin{enumerate}
	
	\item Given the current charge distribution $q(\bm{x})$ and shape of the interface, compute $\llbracket  E^{n}( \bm{x} ) \rrbracket$ by numerically inverting \eqref{eq:BEM:integral_eq_En} using GMRES~\citep{saad1986gmres}. 
	From $\llbracket  E^{n}( \bm{x} ) \rrbracket$, we obtain $E^{{n}+}( \bm{x} )$ and $E^{{n}-}( \bm{x} )$ via \eqref{eq:BEM:q_E_in_out}.
	
	\item Determine the potential $\varphi$ along the interface by evaluating \eqref{eq:BEM:potential_SLP}.
	
	\item Differentiate the surface potential numerically along the interface in order to obtain the tangential electric field $\bm{E}^\mathrm{t}=-\bnabla_{{s}} \varphi$.
	
	\item Knowing both components of the electric field, determine the jump in the electric traction $\llbracket \bm{f}^\mathrm{E}\rrbracket$ and use it to obtain $\llbracket \bm{f}^\mathrm{H}\rrbracket$ using \eqref{eq:BEM:dyn_BC_nondim}.
	
	\item Solve for the interfacial velocity using the Stokes boundary integral equation \eqref{eq:BEM:integral_eq_Stokes}. 
	
	\item Compute $\partial_t q$ via \eqref{eq:pb_def:chgcons} and update the charge distribution using a second-order Runge-Kutta scheme.  
	
	\item Update the position of the interface by advecting the grid with the normal component of the interfacial velocity: $\partial_t{\bm{x}}=(\bm{u}\bcdot\bm{n})\bm{n}$. 
	
	\item Apply corrections to the shape and charge distribution to ensure incompressibility and charge neutrality, as discussed in Sec.~\ref{sec:incompress_charge_neut}. 
	
	\item Reparametrize the interfacial shape following the method discussed in Sec.~\ref{sec:reparam} to minimize high-frequency components in the spherical harmonic expansion. 
\end{enumerate}

\subsection{Surface representation\label{sec:surf_repres}}

The shape of the drop is assumed to be smooth and of spherical topology. Therefore, the surface is parameterized by a truncated series of spherical harmonic expansion from a rectangular domain $\mathbb{S}^2=\{ (\theta,\,\phi)|  \, \theta \in(0,\,\pi), \, \phi \in [0,\, 2\pi)\}$ to $\mathbb{R}^3$: 
\beq \label{eq:SH_x}
\bm{x}(\theta,\phi)=\sum_{n=0}^{N-1}\sum_{m=0}^{n}\, {\bar{P}}_{n}^{m}(\cos \theta) \left( \bm{a}_{nm} \,\cos m\phi \, + \bm{b}_{nm}\,\sin  m\phi \right),
\eeq
where $\theta$ and $\phi$ are the latitude and longitude angles, and $\bm{s}=\{\bm{a}_{nm} ,\, \bm{b}_{nm} \}$ are the coefficients of the expansion in a compact form \citep{boyd2001chebyshev}. The representation above yields $N^2$ spherical harmonic modes per each component of $\bm{x}$ (total of $3N^2$ modes). The normalized associated Legendre polynomials of degree $n$ $(n=0,\, 1,\, 2,\, \dots)$ and order $m$ $(m\leq n)$ are defined as
\beq\label{eq:Pmn_def}
{\bar{P}}_{n}^{m}( \eta)=\frac{1}{2^n \, n!} \sqrt{ \frac{(2n+1) \, (n-m)!} {2 \,(n+m)!}} \, (1-\eta^2) ^{m/2} \, \frac{ \mathrm{d}^{n+m} }{\mathrm{d}x^{\, n+m}}(\eta^2-1)^n,
\eeq
and satisfy the orthogonality condition
\beq
\int_{-1}^{1} \bar{P}^{m}_{n}(\eta) \, \bar{P}^{m}_{n'}(\eta)\, \mathrm{d}\eta= \delta_{nn'}.  \eeq\label{eq:Pmn_orthogonality} Similarly, the surface charge distribution is represented as 
\beq \label{eq:SH_q}
q(\theta,\phi)=\sum_{n=0}^{N-1}\sum_{m=0}^{n}\, {\bar{P}}_{n}^{m}(\cos \theta) \left( \tilde{a}_{nm} \, \cos m\phi \, + \tilde{b}_{nm}\, \sin  m\phi \right).
\eeq
The rectangular domain $\mathbb{S}^2$ is discretized based on the roots of the Legendre polynomial $P_N(\cos \theta)$ along $\theta$, and uniformly along $\phi$.
Forward and backward transformations are performed using the SPHEREPACK library \citep{adams1999spherepack, swarztrauber2000generalized}.
Partial derivatives of a given distribution can be computed using recurrence relations for the derivatives of the associated Legendre polynomials \citep{adams1999spherepack, rahimian2015boundary}.
Besides the spectral accuracy, the spherical harmonic representation allows for nondissipative dealiasing, which improves the numerical stability of the simulations \citep{zhao_freund2010JCP}.

A local coordinate system is constructed at every point $\bm{x}(\theta,\, \phi)$ on the surface of the drop, using two tangent vectors $\bm{a}_{1,2}$ and the unit normal $\bm{a}_3$:
\beq\label{eq:a123}
\centering
\bm{a}_{1}=\partial_\theta \bm{x}, \qquad \bm{a}_{2}=\partial_ \phi \bm{x}, \qquad \bm{a}_3=\bm{n}=\dfrac{\bm{a}_1 \times \bm{a}_2}{|\bm{a}_1 \times \bm{a}_2|}.
\eeq
Consequently, the first and second fundamental forms of the drop surface have the following components:
\beq\label{eq:1st_2nd_fund}
L_{ij}=\bm{a}_i\bcdot\bm{a}_j, \quad \text{and} \quad B_{ij}=\bm{a}_{i,j}\bcdot\bm{n}, \quad (i,j=1,2),
\eeq
which will be used in the subsequent derivations. Given a scalar function $f(\theta,\, \phi)$ on the surface $D$ defined by the parameterization introduced in \eqref{eq:SH_x}, the surface gradient $\bnabla_{s} f$ is 
\beq\label{eq:surf_grad}
\bnabla_{s} \,f = \left(\dfrac{L_{22}\, \bm{a}_1 - L_{12}\,\bm{a}_2}{W^2} \right) 
\partial_\theta f +\, \left( \dfrac{L_{11}\,\bm{a}_2 - L_{12}\,\bm{a}_1}{W^2} \right) 
\partial_\phi f,
\eeq
where $W=(\det{\bm{L}})^{1/2}$ is the area element. Similarly, the surface divergence of a vector field $\bm{v}(\theta,\, \phi)$ can be expressed as
\beq\label{eq:surf_diverg}
\bnabla_{s} \bcdot \bm{v} = \left(\dfrac{L_{22}\, \bm{a}_1 - L_{12}\,\bm{a}_2}{W^2} \right) 
\bcdot \partial_\theta  \bm{v} +\, 
\left( \dfrac{L_{11}\,\bm{a}_2 - L_{12}\,\bm{a}_1}{W^2} \right) 
\bcdot \partial_\phi \bm{v}.
\eeq
In this study, we consider a drop of an incompressible fluid. 
In the absence of Marangoni effects, the capillary stress is a function of the mean curvature $H$:
\beq\label{eq:mean_curv}
H=\frac{1}{2} \Tr(\bm{L}^{-1}\bm{B})= \frac{1}{2}\left(\dfrac{L_{22}\, B_{11}  - 2L_{12}\, B_{12} + L_{11}\,B_{22} }{W^2}\right).
\eeq

\subsection{Aliasing errors\label{sec:dealias}}

Samples of different functions may become indistinguishable on a discrete grid by so-called aliasing~\citep{canuto2012spectral}. 
This means, for instance, that a high-frequency spherical harmonic basis function may be aliased to lower frequencies and cause numerical instability in the simulations. 
Nonlinear operations and differentiation broaden the deformation and charge distribution spectra on a drop. 
As a result, the energy is moved to frequencies not resolved by the grid resolution and may alias to the resolved frequencies~\citep{canuto2012spectral}. 
Therefore, it is physically consistent with the discretization level to remove the corresponding energy from the solution.

\begin{figure}[t]	
	\centering
	\adjustbox{trim={0.0\width} {0.0\height} {0.0\width} {0.0\height},clip}
	{\includegraphics[width=0.6
		\textwidth, angle=0]{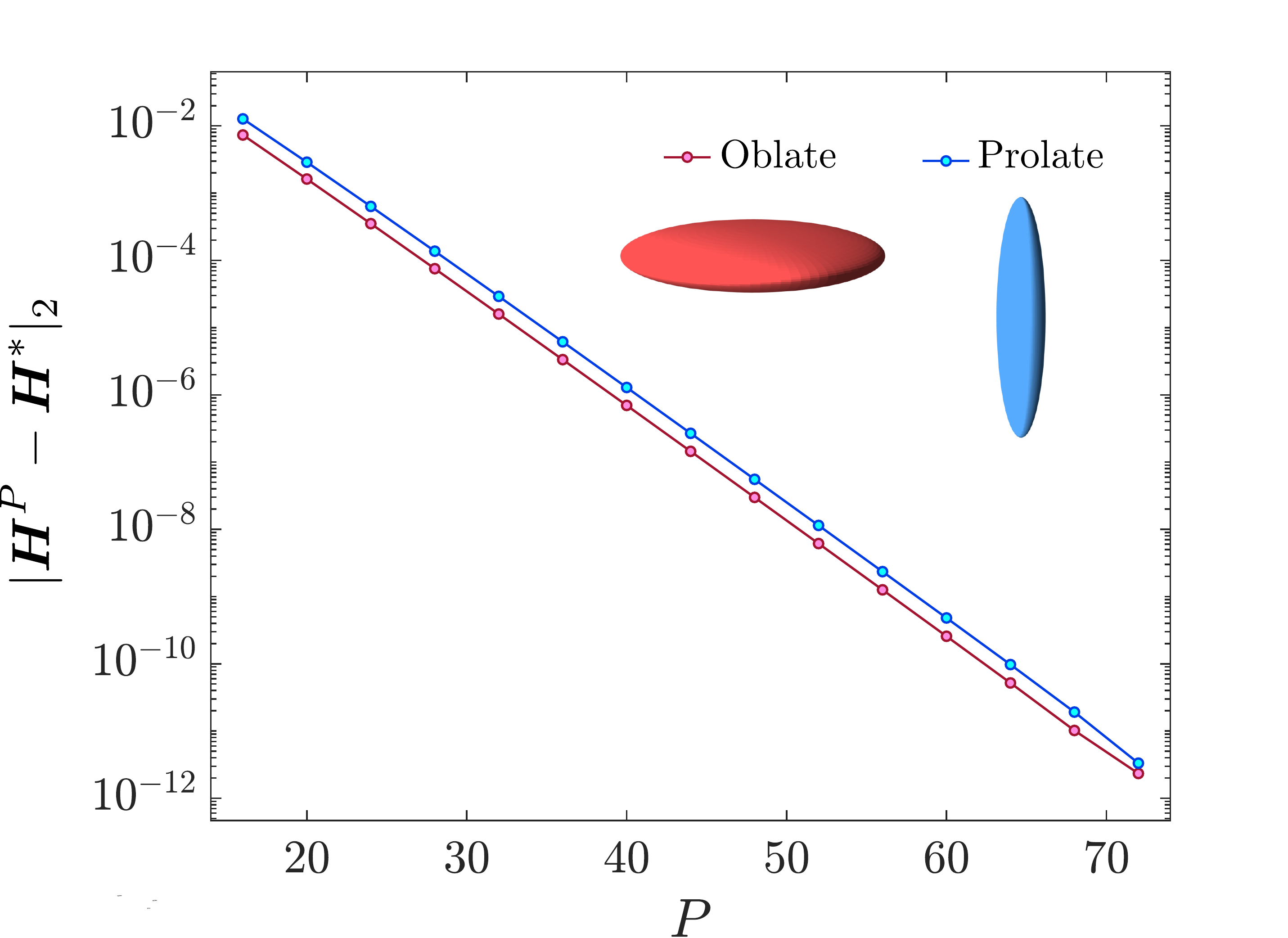} }	
	\caption{ 
		Error in the mean curvature as a function of the number of modes used in the upsampled $P-$space. 
		Two ellipsoids with $l/b=5$ (prolate) and  $l/b=0.2$ (oblate) are investigated, representing the typical shapes of a drop due to EHD flows with large deformations.
		$\boldsymbol{H}^*$ is the mean curvature computed on the finest upsampled grid and is chosen as a surrogate for the exact curvature values.
	}
	\label{figs:curv_decay}  
\end{figure}

Nonlinear operations that are susceptible to aliasing in our simulations include the calculation of mean curvature, electrical stress, convective charge flux, and quadrature.
We use a mesh with finer resolution $M > N $ in the simulations to prevent aliasing errors and filter the solution following any nonlinear manipulation~\citep{zhao_freund2010JCP}.
Our numerical experiments show that $M/N=2$ is sufficient for lower-order and polynomial nonlinearities.
For higher-order and non-polynomial nonlinear manipulations such as those incurred in calculating mean curvature and charge convection, we use an adaptive algorithm that follows  \citet{rahimian2015boundary}.
Given a function $\bm{f}^M$ sampled over an $M$-grid ($2M^2$ points in physical space and $M^2$ spherical harmonic modes), one can interpolate the distribution on a finer $P$-grid $(P/M=u_{f} > 1 )$ by upsampling:
\[
\bm{s}^P=\{\bm{a}_{nm} ,\, \bm{b}_{nm} \}^P= 
\begin{dcases}\label{eq:upsample}
	~\{\bm{a}_{nm} ,\, \bm{b}_{nm} \}^M,& \quad \text{for }~~ n\le M ~~\text{and} ~~ 0\le m \le n,\\
	~0,              & \quad \text{for }~~ n> M.
\end{dcases}
\]
The upsampling factor $u_{f}$ can be determined based on the desired tolerance.
Fig.~\ref{figs:curv_decay} shows how the error in the mean curvature decays as a function of the finer grid resolution $P$ for oblate and prolate spheroids representing typical drop shapes subject to an electric field.

\subsection{Numerical integration\label{sec:integration}}

This section summarizes the numerical integration scheme used in this study, which follows that used by \citet{zhao_freund2010JCP}.
The boundary integrals in \eqref{eq:BEM:potential_SLP}, \eqref{eq:BEM:integral_eq_En}, and \eqref{eq:BEM:integral_eq_Stokes} can be written in the general form
\beq \label{eq:num_integration_I}
I(\bm{x}_0) = \int_D K(\bm{x}, \bm{x}_0)\, f(\bm{x})\, \mathrm{d}s(\bm{x}) = \int_{\mathbb{S}^2} K(\bm{x}(\theta,\, \phi), \bm{x}_0)\, f(\bm{x}(\theta, \phi) ) \, J(\bm{x}(\theta, \phi)) \, \sin \theta\, \mathrm{d}\theta \, \mathrm{d}\phi,
\eeq
where $f(\theta, \phi)$ is a smooth scalar function over surface $D$, and $K$ is the kernel containing the Green's function for Stokes or Laplace's equations.

The approximate spacing of a mesh with $N\times 2N$ points is $h=\sqrt{A/2N^2}$, where $A$ is the drop surface area.
Due to the singular behavior of $K$ as $\bm{x}_0 \rightarrow \bm{x}$,  the drop surface $D$ is divided into two regions where the integral $I$ has different behaviors.
Following the method of floating partition of unity \citep{bruno2001FPOU_integration, ying2006polarpatch, zhao_freund2010JCP}, a local polar patch is considered centered around $\bm{x}_0$.
For any point $\bm{ x}$ on $D$, $\rho(\bm{x}, \bm{x}_0)$ is defined as the distance along the great circle that connects $\bm{x}$ to $\bm{x}_0$ on $\mathbb{S}^2$.
In the next step, a mask function is defined based on this coordinate as
\[
\eta(\rho) =
\begin{dcases}\label{eq:num_integration_eta}
	~ \exp{\left( \dfrac{2e^{-1/t}}{t-1} \right)},& \quad \text{for }~~  t=\rho/\rho_1 <1,\\
	~ 0 & \quad \text{for }~~ \rho\ge \rho_1,
\end{dcases}
\]
where $\rho_1$ is the cutoff radius.
This representation allows for accurate calculation of the integral \eqref{eq:num_integration_I} as $\eta$ is a smooth function.

Next, the surface integral \eqref{eq:num_integration_I} is split into two parts:
\beq\label{eq:num_integration_I1_I2}
I = I_1 +I_2 = \int_D K(\bm{x}, \bm{x}_0)\, 
\eta( \rho(\bm{x}, \bm{x}_0) )\, 
f(\bm{x})\, \mathrm{d}s(\bm{x}) + 
\int_D K(\bm{x}, \bm{x}_0)\, [1-\eta( \rho(\bm{x}, \bm{x}_0) )]\, f(\bm{x})\, \mathrm{d}s(\bm{x}).
\eeq
The integrand of $I_1$ has support only inside the patch.

Using a transformation to the local polar coordinate system $(\rho,\varphi)$, we write:
\beq\label{eq:num_integration_I1_rho_phi} I_1=\int_0^{2 \pi} \int_D^{\rho_1} K(\bm{x},  \bm{x}_0)\, \eta(\rho)f(\rho, \varphi) \sin{\rho}\, \mathrm{d} \rho \, \mathrm{d} \varphi, \eeq
where the integrand is finite and periodic in $\varphi$.
The integral \eqref{eq:num_integration_I1_rho_phi} is computed using Gauss quadrature along $\rho$ from $0$ to $\rho_1$, with a uniform mesh in $\varphi$ from $0$ to $2\pi$.
We set the patch radius on the reference sphere to $\rho_1=\pi/\sqrt{N}$, which means the patch radius is $O(h^{1/2})$ in $\mathbb{S}^3$.
Inside the patch, $\sqrt{N}$ points are considered along $\rho$, and $2\sqrt{N}$ points along $\varphi$.
We note that the quadrature points inside the patch do not coincide with the surface mesh points.
Therefore, bi-cubic spline interpolation is used to evaluate the coordinates and other functions at quadrature points of $I_1$.
The error of singular integration with the mentioned choice of patch size is $O(h^3)$~\citep{ying2006polarpatch}.
Higher-order accuracy can be achieved using a larger patch size at the expense of computational cost.

The second part of integral \eqref{eq:num_integration_I1_I2} is $I_2$, which has a smooth integrand. Therefore, it is computed accurately as:
\beq\label{eq:num_integration_I2_quadrat}
I_2 \approx \sum_{i=1}^{N}\sum_{j=1}^{2N} K(\bm{x}_{ij}, \bm{x}_0) \,\eta( \rho(\bm{x}_{ij}, \bm{x}_0) )\, f_{ij}\, J_{ij}\, \mathit{w}_{ij},
\eeq
where $\bm{x}_{ij}=\bm{x}(\theta_i,\, \phi_j)$ are the quadrature points, $w_{ij}$ are the corresponding weights and $J_{ij}$ is the Jacobian of the transformation from $\mathbb{S}^2$ to $\mathbb{R}^3$.
The quadrature \eqref{eq:num_integration_I2_quadrat} converges exponentially with the mesh size $h$, using quadrature points with Gaussian and uniform distributions along the $\theta$ and $\phi$ directions, respectively.
Further details on the numerical integration scheme can be found in \citet{zhao_freund2010JCP}.

\subsection{Incompressibility and charge neutrality\label{sec:incompress_charge_neut}}

We expect no change in the drop volume, as both fluid phases are incompressible.
However, small changes in the volume occur due to numerical errors, which we correct by adjusting the shape of the drop along the normal direction~\citep{zhao_freund2010JCP}.
In the reported simulations, the magnitude of these adjustments is smaller than $10^{-8}r_0$ at every time step, where $r_0$ is the initial drop radius.
Similarly, the net charge (less than $~10^{-17}\varepsilon E_0$) is subtracted from the surface charge distribution at every time step to ensure charge neutrality throughout the simulations.
The mentioned corrections are especially important for long simulation times.

\subsection{Reparametrization\label{sec:reparam}}

The fluid-fluid interface of the drop evolves and deforms. 
During this process, there is no physical mechanism to inhibit in-plane distortions of the grid (i.e., depletion, aggregation, and skewness) since there is no bending rigidity, in-plane shear resistance, or surface inextensibility, 
High-frequency components of the spherical harmonic expansions thus grow, exacerbating aliasing errors and resulting in numerical instability.
In addition to correcting aliasing errors, as explained in Sec.~\ref{sec:dealias}, one must develop a reparameterization strategy that ensures stable and accurate simulations over long simulation times.
Here, we use an algorithm that minimizes the high-frequency components in the spherical harmonic expansion of the surface parametrization.
This method was introduced by Veerapenani et al.\ \citep{veerapaneni2011fast, rahimian2015boundary} and improved by \citet{sorgentone2018surfactant_JCP}.

Consider an implicit representation of the interface as a smooth function $F:\mathbb{R}^3\mapsto \mathbb{R}$ such that $F(\bm{x})=0$ for all $\bm{x} \in D$ where $D$ is the drop surface.
The unit normal vector can be expressed as $\bm{n}=\bnabla F/ |\bnabla F|$ at every point on the surface.
We define a quality metric $E:\mathcal{X}\mapsto \mathbb{R}$ where $\mathcal{X}$ is the space of smooth functions defined over $D$.
Now, we can view the reparameterization problem as a minimization of $E(\bm{x})$ subject to the constraint $F(\bm{x})=0$ as
\beq
\min_{\bm{x}\in D}\{ E(\bm{x})\}~\text{subject to}~  F(\bm{x})=0.
\eeq
It can be shown that the solution to the problem above is
\beq\label{eq:Energy_minmization}
(\bm{I}-\bm{n}(\bm{x}) \bm{n}(\bm{x}) )\bcdot \bnabla E(\bm{x})=\bm{0}, \quad \text{and} ~~F(\bm{x})=0.
\eeq
The choice of quality metric $E(\bm{x})$ is not unique and could vary depending on the reparametrization strategy.
\cite{veerapaneni2011fast} proposed an equivalent of the following quality metric:
\beq\label{eq:spectral_energy}
E(\bm{x})=\sum_{n=0}^{N-1}\sum_{m=0}^{n} \alpha _{nm} \left( |\bm{a}_{nm}|^2_{2}+|\bm{b}_{nm}|^2_{2} \right),
\eeq
where $|\bm{y}|_2$ is the $L^2$ norm of vector $\bm{y}$, and $\alpha_{nm}$ is the weight for $(n,m)$-th spherical harmonic.
$E$, as defined in \eqref{eq:spectral_energy}, can be viewed as the spectral energy of the parametrization.
We aim to minimize the high-frequency part of $\bm{x}$.
Therefore, the weights $\alpha_{nm}$ must be small for low frequencies and larger for high frequencies.
Here, we use the following perfect low-pass filter
\[
\alpha_{nm}= 
\begin{dcases}\label{eq:low-pass-filter}
	1,& \quad \text{for }~~ n> N_{\text{cutoff}},\\
	~0,              & \quad \text{for }~~ n\le N_{\text{cutoff}}.
\end{dcases}
\]
that was used \citep{veerapaneni2011fast,sorgentone2018surfactant_JCP} to simulate vesicles and surfactant-laden drops and improved long-time simulations.
$N_{\text{cutoff}}$ is the cutoff frequency and is chosen adaptively based on the energy spectrum as
\beq\label{eq:N_cutoff}
N_{\text{cutoff}} = \min \{k\in \mathbb{N}, ~~1\le k \le N-1~~| ~~ E_k/E_1\le P_{\text{cutoff}}  \},
\eeq
where $E_k= \sum_{n=k}^{N-1}\sum_{m=0}^{n} \alpha _{nm} ( |\bm{a}_{nm}|^2_{2}+|\bm{b}_{nm}|^2_{2} )$ and $P_{\text{cutoff}}$ determines the fraction of modes we penalize ($P_{\text{cutoff}}=0.2$ in our simulations) \citep{sorgentone2018surfactant_JCP}. 

We solve \eqref{eq:Energy_minmization} by marching in pseudo-time $\tau$ along the tangential direction at every point:
\beq \label{eq:pseudo_velocity_x_tau}
\partial_\tau \bm{x}+(\bm{I}-\bm{n}(\bm{x}) \bm{n}(\bm{x}) ) \bcdot \bnabla E(\bm{x})=\bm{0},\quad \mathrm{with}\,\,\,\,\bm{x}( \tau=0)=\bm{x}_0.
\eeq
However, following \eqref{eq:pseudo_velocity_x_tau} the volume is not necessarily conserved.
In addition, the charge distribution is distorted and hence not spectrally accurate.
We use the method of \citet{sorgentone2018surfactant_JCP} to address these challenges.
We project the linear pseudo-velocity $\partial_\tau \bm{x}$ along the latitudinal and longitudinal directions to obtain the angular pseudo-velocity $(\partial_\tau \theta,\, \partial_\tau \phi)$.
This strategy comprises updates of the angular coordinates $(\theta,\, \phi)$ for every point and uses the original spherical harmonic expansions for the shape and charge distribution to interpolate the updated distributions.
Using the chain rule, 
\beq
\partial_\tau \bm{x} = \frac{\partial \bm{x}}{\partial \theta} \,\frac{\partial \theta}{ \partial \tau} + \frac{\partial \bm{x}}{\partial \phi} \,\frac{\partial \phi}{ \partial \tau}= \bm{a}_1 \,\partial_\tau \theta + \bm{a}_2 \, \partial_\tau \phi.
\eeq
Next, we project  $\partial_\tau \bm{x}$ along the tangential directions by taking inner products with $\bm{a}_1$ and $\bm{a}_2$:
\begin{align}\label{eq:xt_tagent_eq_sys1}
	& \partial_\tau \bm{x}\bcdot \bm{a}_1=  ~~ |\bm{a}_1|^2 \, ~~\partial_\tau \theta +  \bm{a}_1\bcdot \bm{a}_2 \,~ \partial_\tau \phi,\\\label{eq:xt_tagent_eq_sys2}
	& \partial_\tau \bm{x}\bcdot \bm{a}_2=   \bm{a}_1\bcdot \bm{a}_2 \,~ \partial_\tau \theta +  ~~|\bm{a}_2|^2 \, ~~\partial_\tau \phi.
\end{align}
The solution to \eqref{eq:xt_tagent_eq_sys1} and \eqref{eq:xt_tagent_eq_sys2} is
\beq
\partial_\tau \bm{\theta} = \bm{L}^{-1}\bm{w},
\eeq
where $\partial_\tau \bm{\theta}=\{ \partial_\tau \theta,\, \partial_\tau \phi \}^T$ and $\bm{w} = \{ \partial_\tau \bm{x}\bcdot \bm{a}_1, \, \partial_\tau \bm{x}\bcdot \bm{a}_2\}^T$,  and $\bm{L}$ is the surface metric defined in \eqref{eq:1st_2nd_fund}.
We update the angular coordinates of every grid point by marching explicitly in pseudo-time to obtain new angles $(\theta^*,\, \phi^*)$.
Using the new angular coordinates, we interpolate the position and surface charge at every point based on the spherical harmonic expansions of the grid before reparametrization (at $\tau=0$): 
\begin{align}
	\bm{x}^*(\theta^*,\phi^*)=\sum_{n=0}^{N-1}\sum_{m=0}^{n}\, {\bar{P}}_{n}^{m}(\cos \theta^*) \left( \bm{a}^0_{nm} \,\cos m\phi^* \, + \bm{b}^0_{nm}\,\sin  m\phi^* \right),\\
	q^*(\theta^*,\phi^*)=\sum_{n=0}^{N-1}\sum_{m=0}^{n}\, {\bar{P}}_{n}^{m}(\cos \theta^*) \left( \tilde{a}^0_{nm} \, \cos m\phi^* \, + \tilde{b}^0_{nm}\, \sin  m\phi^* \right).
\end{align}
Finally, a forward (FSHT) and backward (BSHT) spherical harmonic transformation of $\bm{x}^*$ and $q^* $ yields distributions on a standard grid that is spaced uniformly along $\phi$ and Gaussian along $\theta$:
\begin{align}
	& \bm{x}^*\, \xrightarrow{\text{FSHT}}\, \{\bm{a}_{nm},\, \bm{b}_{nm}\}   \xrightarrow{\text{BSHT}}\, \bm{x},\\
	& q^*\, \xrightarrow{\text{FSHT}}\, \{\tilde{a}_{nm},\, \tilde{b}_{nm}\} \xrightarrow{\text{BSHT}}\, q.
\end{align}

\section{Numerical results\label{sec:results}}

In this section, we compare the results of our numerical method with existing theoretical solutions and previous computational and experimental studies on electrohydrodynamic flows in drops.
Table~\ref{table:list_systems} lists all systems considered in this study, including their dimensional and non-dimensional parameters.

\begin{table}[t] 
	\centering
	\caption{
		Physical systems studied using numerical simulations, with their dimensional and non-dimensional parameters. 
		$\mu^+ = {0.69}\, {\mathrm{Pa\, s} }$, $ \gamma = {4.5} \,\mathrm{mN \,m^{-1}}$,  and $ \epsilon^0 = {8.8542e-12}\, \mathrm{F\, m^{-1}}$ denotes the permittivity of vacuum.\vspace{0.2cm}
		\label{table:list_systems}}
	{\small
		\begin{tabular}{c c c c c c | c c c}
			system & $\mathrm{R}$ & $\mathrm{Q}$ & $\mathrm{\lambda}$ & $\mathrm{Ca}_\mathrm{E}$ & $\mathrm{Ma}$ & $\sigma^+~(\mathrm{S\, m^{-1}})$ & $\epsilon^+/\epsilon^0$  &$r_0~(\mathrm{mm})$ \\[0.5ex] 
			\hline \hline
			$\mathbi{S1}$   & $0.57$ & $36.59$ & $1.41$ & $0.05-0.75$ & $0.21-3.25~$ & $4.5\times 10^{-11}$ & $5.3$ &    $0.7$  \\
			$\mathbi{S2}$  & $0.57$ & $36.59$ & $1.41$ & $0.05-0.75$ & $0.65-9.75~$ & $4.5\times 10^{-11}$ & $5.3$ &   $2.1$ \\
			$\mathbi{S3}$  & $0.57$ & $36.59$ & $14.12$ & $0.01-1.44$ & $0.28-5.0~$ & $4.5\times 10^{-11}$ & $5.3$ &   $0.25,~0.75,~1.25,~1.75$ \\
			$\mathbi{S4}$  & $0.60$ & $10.56$ & $0.07$ & $0.15-1.0$ & $0.43-2.86$ & $3.8\times 10^{-11}$ & $4.7$ &  $1.9$ \\
		\end{tabular}
	}
\end{table}

\subsection{Axisymmetric regime\label{sec:results:taylor_regime}}

\begin{figure}[t]	
	\centering
	\adjustbox{trim={0.0\width} {0.0\height} {0.0\width} {0.0\height},clip}
	{\includegraphics[width=1.
		\textwidth, angle=0]{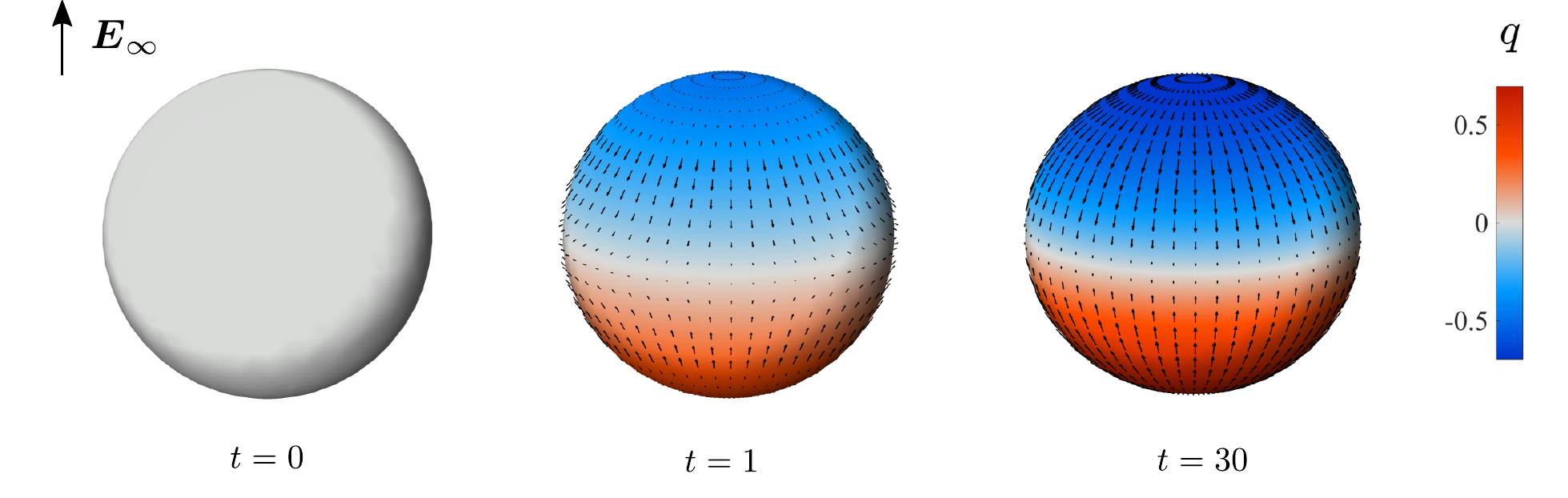} }	
	\caption{Axisymmetric regime: evolution of a drop of system $\mathbi{S2}$ subject to a uniform electric field with $\mathrm{Ca_{E} } =0.2$ ($N=8,~M=3N$). Colors show the interfacial charge density $q$, while arrows show the interfacial fluid velocity. {Also see the corresponding video in the Supplemental Material. } }  
	\label{figs:validation_Taylor_regime_flow}  
\end{figure}

\begin{figure}[h]	
	\centering
	\adjustbox{trim={0.0\width} {0.0\height} {0.0\width} {0.0\height},clip}
	{\includegraphics[width=1.0\textwidth, angle=0]{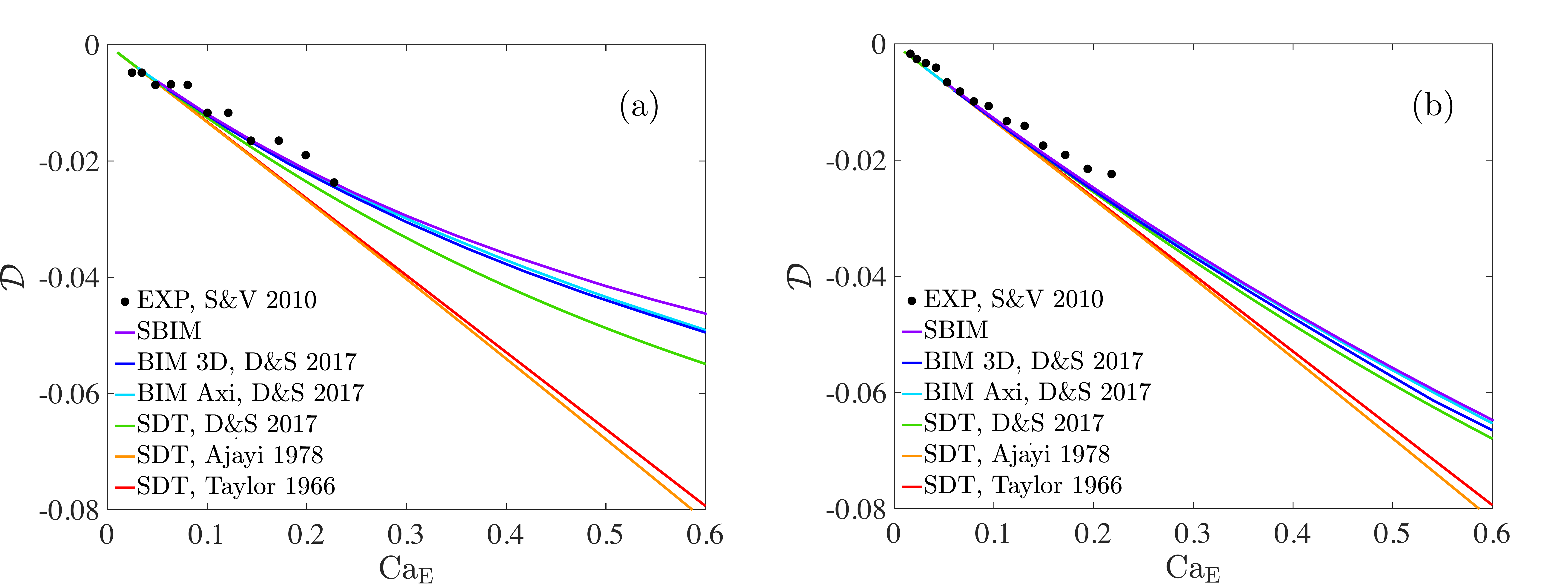} }	
	\caption{
		Steady drop deformation as a function of electric capillary number $\mathrm{Ca}_\mathrm{E}$ for systems $\mathbi{S1}$ in $(a)$ and  $\mathbi{S2}$ in $(b)$. Results from our spectral boundary integral method (SBIM) are compared with the experimental data (EXP) of \citet{salipante_valhovska2010EHD_drop_PRF}, along with axisymmetric and three-dimensional boundary integral simulations (BIM) by Das and Saintillan  \citep{das2017EHD_simulation, das2017nonlinear} and various small-deformation theories (SDT) \citep{taylor1966LDM,rallison1978numerical,das2017nonlinear}. We use $N=8$ and $M=3N$ in the simulations shown here.
	}
	\label{figs:validation_Taylor_regime}  
\end{figure}

Under weak electric fields ($\mathrm{Ca}_\mathrm{E}\ll 1$), the drop adopts a steady axisymmetric shape and charge distribution, as shown in Fig.~\ref{figs:validation_Taylor_regime_flow}. Following the pioneering work of Taylor \citep{taylor1966LDM}, various  analytical models have been proposed to improve the predictions of the steady shape of a drop subject to weak electric fields  and small deformations  \citep{rallison1978numerical,das2017nonlinear}.
We characterize the deviation from the spherical shape using Taylor's deformation parameter $\mathcal{D}$ defined as
\beq
\mathcal{D}=\dfrac{l-b}{l+b},
\eeq
where $l$ and $b$ are the lengths of the major axes of the drop parallel and perpendicular to the applied electric field.
Figure~\ref{figs:validation_Taylor_regime} shows the steady-state deformation $\mathcal{D}$ from our simulations compared to the predictions of different small-deformation theories (SDT), prior numerical simulations, as well as experimental data.
As shown in Fig.~\ref{figs:validation_Taylor_regime}, our simulation results closely match the experimental data and past predictions.
In agreement with past studies \cite{lanauze2015EHD_JFM,das2017nonlinear,das2017EHD_simulation}, interfacial charge convection by the fluid flow plays a significant role in the dynamics and tends to weaken drop deformations.

\subsection{Quincke regime\label{sec:results:quincke}}

\begin{figure}[t]	
	\centering
	\adjustbox{trim={0.0\width} {0.0\height} {0.0\width} {0.0\height},clip}
	{\includegraphics[width=1.0
		\textwidth, angle=0]{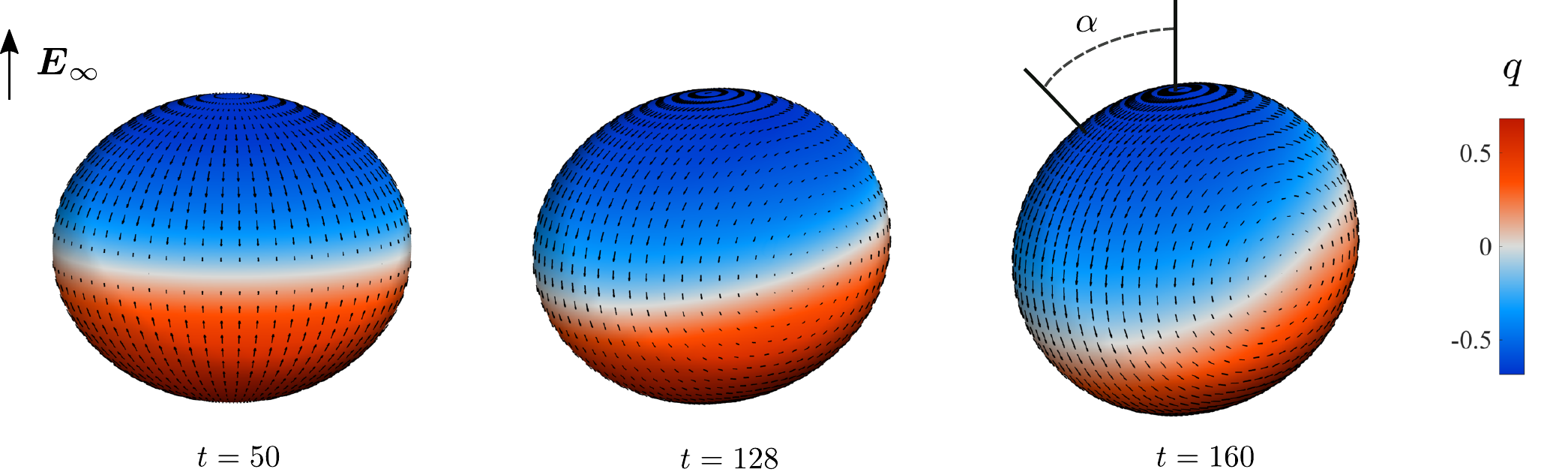} }	
	\caption{
		Quincke regime: spontaneous rotation and tilt of a drop of system $\mathbi{S3}$  subject to an applied electric field with $E_\infty/E_\mathrm{c}=1$ and $\mathrm{Ca}_{\mathrm{MW}} =1.32$ ($N=10,~M=3N$). 
		The tilt angle $\alpha$ is measured from the initial orientation of the drop as depicted. {See the Supplemental Material for the corresponding video showing the evolution of the drop in the Quincke regime. }
	}  
	\label{figs:validation_Quincke_regime}  
\end{figure}

Upon increasing the intensity of the electric field, the drop can undergo a symmetry-breaking bifurcation to a dynamical regime characterized by a rotational component to the EHD flow, as shown in Fig.~\ref{figs:validation_Quincke_regime} \citep{ha2000EHD_quincke,sato2006EHD_quincke, salipante_valhovska2010EHD_drop_PRF}.
When $\mathrm{RQ }>1$, the induced flow is from the poles to the equator, and the induced electric dipole is anti-parallel to the applied electric field.
This condition is unfavorable for stability as perturbations to the system result in a destabilizing electric torque that can drive spontaneous rotation beyond a critical electric field strength $E_\mathrm{c}$.
Quincke first observed a similar phenomenon in solid spheres subject to uniform electric fields \citep{quincke1896ueber}.
In solid spheres, the threshold for Quincke electrorotation is given by \citep{jones1984quincke}
\beq
E_\mathrm{c}=\sqrt{\frac{2\mu^+}{3 \epsilon^+ \,\tau_{\mathrm{MW}}\, (\bar{\epsilon} - \bar{\sigma})}}, \quad \mathrm{where}\quad  \bar{\sigma}=\frac{\sigma^- - \sigma^+}{\sigma^- +2\sigma^+}, \quad \bar{\epsilon}=\frac{\epsilon^- - \epsilon^+}{\epsilon^- +2\epsilon^+}.
\eeq

\begin{figure}[t]	
	\centering
	\adjustbox{trim={0.0\width} {0.0\height} {0.0\width} {0.0\height},clip}
	{\includegraphics[width=1.0
		\textwidth, angle=0]{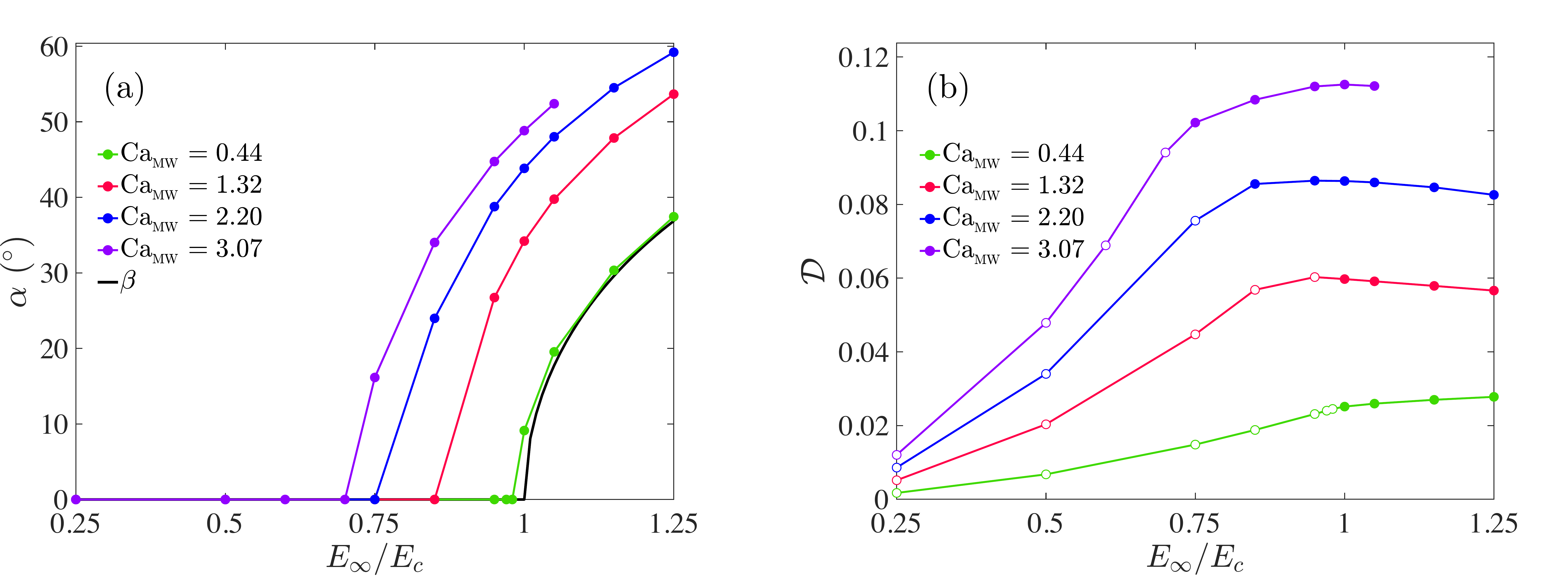} }	
	\caption{
		Steady tilt angle $\alpha$ (a), and deformation $\mathcal{D}$ (b) as functions of the applied electric field $E_{\infty}/E_\mathrm{c}$ for different values of $\mathrm{Ca}_{\mathrm{MW}} $ (different $r_0$) for system ${\mathbi{S3}}$. 
		The solid black line in (a) shows the tilt angle $\beta$ of a rigid sphere. 
		Filled markers in (b) show cases where Quincke electrorotation occurs. 
		We used $N=10$ and $M=3N$ in all simulations. 
	}  
	\label{figs:validation_quincke}  
\end{figure}

Figure \ref{figs:validation_Quincke_regime} shows the velocity field and charge distribution in a leaky dielectric drop during Quincke electrorotation.
Following the onset of instability, the drop tilts away from its initial orientation.
The tilt angle $\alpha$ and deformation oscillate until they reach their steady-state values.
Figure \ref{figs:validation_quincke} shows the steady-state tilt angle and deformation as a function of the applied electric field $E_\infty / E_\mathrm{c}$ for a drop of the system $\mathbi{S3}$.
For a given set of material properties, different $\mathrm{Ca}_{\mathrm{MW}}$ correspond to different drop sizes, and the effect of capillary forces is stronger for smaller drops (small $\mathrm{Ca}_{\mathrm{MW}}$).
Therefore, we expect the drop tilt angle $\alpha$ to approach that of a solid sphere $\beta$ \citep{jones1984quincke,salipante_valhovska2010EHD_drop_PRF}:
\beq
\beta=\frac{\pi}{2}-\arctan\left( \frac{E^2_\infty}{E^2_\mathrm{c}}-1 \right)^{-\frac{1}{2}}, \qquad \text{for} ~~{E_\infty}\ge{E_\mathrm{c}},
\eeq
for sufficiently large $\mathrm{ \lambda }$ and small $\mathrm{Ca}_\mathrm{MW}$.
This is verified in Fig.~\ref{figs:validation_quincke}~(a), where the green curve for $\mathrm{Ca}_\mathrm{MW}=0.44$ closely follows the black line corresponding to $\beta$.

\subsection{Low-viscosity drops\label{sec:results:low_viscosity_drops}}

When a leaky dielectric drop with $\mathrm{RQ}<1$ is subject to a uniform electric field, it compresses along the field direction.
The induced quadrupolar flow is from the poles to the equator at the interface.
Under weak electric fields, the characteristic interfacial velocity is:
\begin{equation}\label{eq:taylor_vel_scale}
	u_{\mathrm{T}}= \dfrac{9}{10} \mathrm{ \dfrac{R(RQ-1)}{(2R+1)^2\,(1+\lambda)} }\, r_{0}\tau^{-1}_{\mathrm{EHD}}
\end{equation}
based on Taylor's classic solution \citep{taylor1966LDM}.
According to Eq.~\eqref{eq:taylor_vel_scale}, we expect the EHD flow to be stronger in low-viscosity drops ($\lambda< 1$) as $u_{\mathrm{T}}\propto (1+\lambda)^{-1}$.
As we increase the intensity of the applied electric field, low-viscosity drops may undergo different types of EHD instabilities, such as dimpling, equatorial streaming, or Quincke rotation, depending on their material properties \citep{brosseau2017streaming}.
In other experiments, colloidal particles adsorbed on a drop interface were observed to accumulate at the equator and form a belt,
which broke into vortices of particles \citep{dommersnes2013colloids_drop, ouriemi2014equatorial_vortices}.
In this section, we study the dynamics of an oblate drop with a small viscosity ratio  $\lambda<0.1$, represented by $\mathbi{S4}$, which is similar to the systems studied in \citep{ouriemi2014equatorial_vortices}.

\begin{figure}[t]	
	\centering
	\adjustbox{trim={0.0\width} {0.0\height} {0.0\width} {0.0\height},clip}
	{\includegraphics[width=1.
		\textwidth, angle=0]{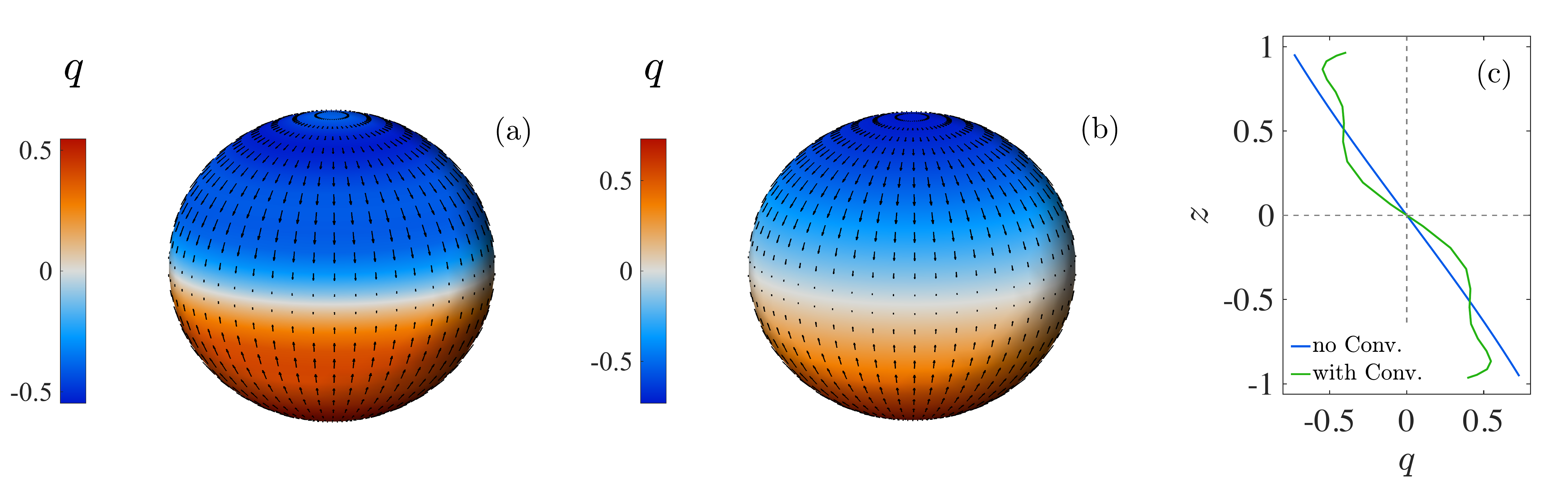} }	
	\caption{
		Charge dynamics in a low-viscosity drop of $\mathbi{S4}$ with $(\mathrm{Ca}_\mathrm{E},\, \mathrm{Ma}) =(0.3,\,1.43)$: steady-state velocity and charge density fields in the presence of charge convection (a), and with no convection (b), and the corresponding charge profiles along the direction of the applied electric field (c). We used $N=8,~M=3N$ in both simulations.
	}  
	\label{figs:ca03_ringing}  
\end{figure}

Our results show that charge convection significantly affects the system's dynamics in this regime.
As a result of the nonlinear coupling between the flow and charge dynamics, strong charge gradients build up around the equator, as shown in Fig.~\ref{figs:ca03_ringing}.
The nonlinear steepening disappears when we switch off the convective term in the charge conservation equation \eqref{eq:pb_def:chgcons}.
Therefore, we can conclude that the charge convection is responsible for the nonlinear behavior mentioned.
This can be verified in Fig.~\ref{figs:ca03_ringing}(c) by comparing the charge profiles for simulations with and without convection.
The convergent EHD flow sweeps charges of opposite signs towards the equator, which is the stagnation line of the flow where the electric field is locally tangent to the interface.
It has been shown that in systems where an interface is subject to a tangential electric field and a converging flow, strong charge gradients develop around the stagnation line due to the effect of  charge convection \citep{firouznia2022JFM}.

The emergence of sharp local features in the charge profile poses a fundamental challenge for our numerical method, as spectral methods are most efficient for smooth field variables.
It is known that large gradients or discontinuities in a function result in Gibbs phenomenon or ringing artifacts in its spherical harmonic representation \citep{gelb1997gibbs_resolution_sh}.
Following the emergence of ringing artifacts, spurious oscillations cannot be contained locally as spherical harmonics are global basis functions.
This phenomenon occurs due to the inability of the finite spherical harmonic expansions to properly represent the infinite (or very large) derivatives in the sharp regions of a discontinuous (or nearly  discontinuous) function.

\begin{figure}[t]	
	\centering
	\adjustbox{trim={0.0\width} {0.0\height} {0.0\width} {0.0\height},clip}
	{\includegraphics[width=1.
		\textwidth, angle=0]{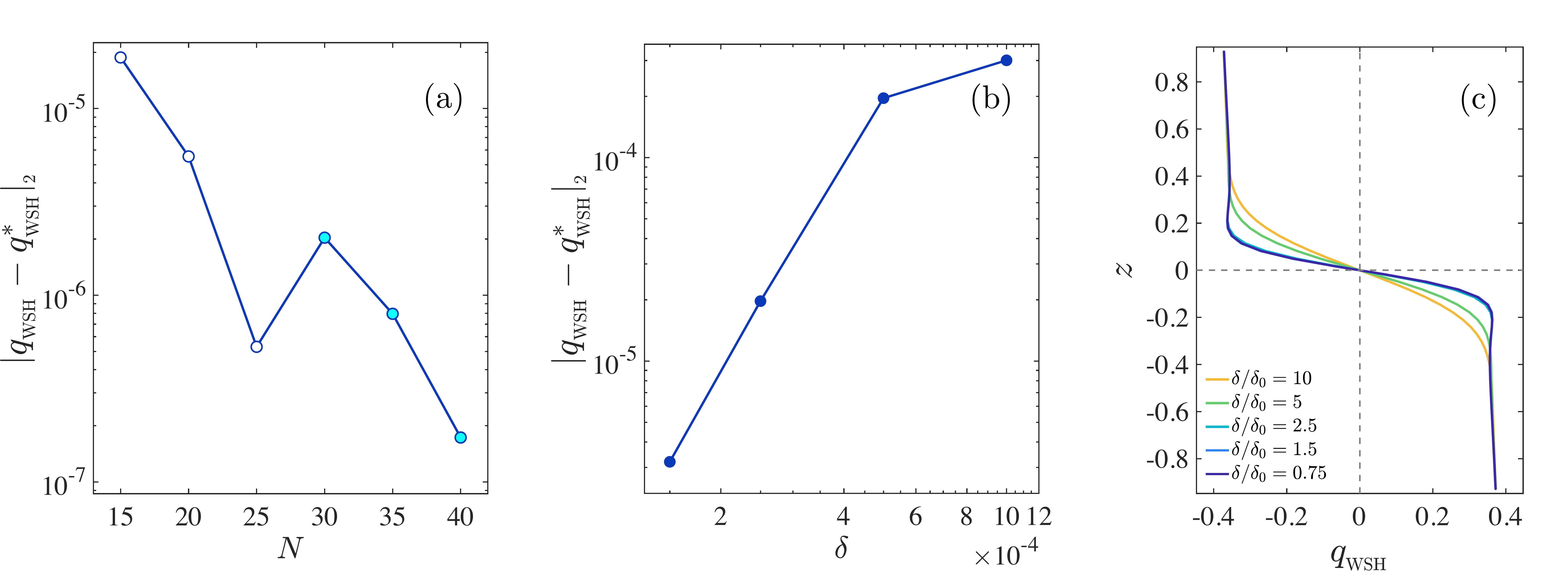} }	
	\caption{
		WSH convergence study on a drop of $\mathbi{S4}$ with $(\mathrm{Ca_E},\, \mathrm{Ma}) =(1.0,\,0.43)$: (a) Error in the steady-state charge density distribution as a function of mesh resolution $N$ for a fixed value of $\delta=10^{-3}$ ($N^*=42$).
		Open markers correspond to cases with ringing artifacts.
		(b) Error in the steady-state charge density distribution as a function of relaxation factor $\delta$ ($N^*=80$, $\delta/\delta_0=0.75$ and $\delta_0=10^{-4}$ for $\smash{q^*_{\mathrm{WSH}}}$).
		(c) Interfacial charge profiles corresponding to (b). {See the simulation video for $N^*=80$, $\delta/\delta_0=0.75$ and $\delta_0=10^{-4}$ in the Supplemental Material. }
	}
	\label{figs:convergence_N_delta}  
\end{figure}

For relatively small electric capillary numbers $\mathrm{Ca_E}<0.2$, we can suppress ringing artifacts by using a finer mesh corresponding to a larger number of spherical harmonics modes, a more restricted time step size, and, therefore, a higher computational cost.
However, due to the stronger effect of charge convection at larger values of $\mathrm{Ca_E}$, refining the resolution is insufficient to remove the ringing artifacts when $\mathrm{Ca_E}$ is increased.

To resolve this problem, we propose a weighted spherical harmonic expansion (WSH) for the charge density distribution, where the contribution from high-frequency components is exponentially relaxed \citep{chung2005heatkernel_smoothing}.
For a given charge density distribution $q(\theta,\phi)$, the spherical harmonic expansion \eqref{eq:SH_q} gives coefficients $\tilde{s}=\{ \tilde{a}_{nm} ,\, \tilde{b}_{nm} \}$ where $n=0,\, 1,\, 2,\, \dots, \, N-1$ and $m\leq n$.
Accordingly, we define $\text{WSH}\!:q\mapsto q_{\mathrm{WSH}}$ as:
\beq \label{eq:WSH_q}
q_{\mathrm{WSH}}(\theta,\phi)=\sum_{n=0}^{N-1}\sum_{m=0}^{n}\, e^{-n(n+1)\delta}\left( \tilde{a}_{nm} \, \cos m\phi \, + \tilde{b}_{nm}\, \sin  m\phi \right) \,{\bar{P}}_{n}^{m}(\cos \theta) ,
\eeq
where $\delta$ is the relaxation factor.
Based on \eqref{eq:WSH_q}, we note that WSH does not affect the mean, and $q_{\mathrm{WSH}}$ converges to $q$ in the limit of $\delta \rightarrow 0 $.
In Fig.~\ref{figs:convergence_N_delta}(a), we demonstrate that increasing the mesh resolution at a fixed value of $\delta$ eliminates the ringing artifacts.
In addition, Fig.~\ref{figs:convergence_N_delta}(b) shows that the steady-state charge profile at a fixed resolution converges as the relaxation factor $\delta$ decreases.
{We combine both approaches and increase $N$ while decreasing $\delta$ to refine the numerical solution while avoiding the emergence of ringing artifacts.
	The numerical solution with the finest resolution, $N^*=\text{max}(N)$, is then chosen as a surrogate to the exact solution in Fig.~\ref{figs:convergence_N_delta}(a,b).
	To compare the charge profiles computed on different mesh resolutions, we upsample them to a fine grid with $N^{u} > N^*$.
	Convergence of the charge profile is illustrated in Fig.~\ref{figs:convergence_N_delta}(c), where the coupling of the flow and charge dynamics is found to result in the formation of strong gradients in the charge density profile near the equator, consistent with previous studies \cite{das2017nonlinear}.
	We note that our simulations for system ${\mathbi{S4}}$ under electric capillary numbers of up to $\mathrm{Ca_E} \sim  O(1)$ reach a steady axisymmetric profile and do not exhibit any electrohydrodynamic instability, suggesting that the equatorial vortices observed by \citet{ouriemi2014equatorial_vortices} could be a consequence of the colloidal particles present at the interface in their experiments.}

\section{Conclusions\label{sec:conclusions}}

We developed a spectral boundary integral method for simulating electrohydrodynamic flows in leaky dielectric viscous drops.
The drop surface, charge density, and all other variables are represented using truncated series of spherical harmonic expansions.
In addition to the excellent accuracy of the spectral representation, it enables us to develop a nondissipative dealiasing method required for numerical stability.
The charge transport is modeled using a conservation equation for Ohmic conduction from the bulk and surface charge convection by the flow and finite charge relaxation.
We employ a reparametrization technique which allows us to explore regimes where drops undergo significant deformations.
Our results closely match the existing experimental data and analytical predictions in the axisymmetric Taylor and Quincke electrorotation regimes.{{The code used in this article is publicly available at \url{https://github.com/mfirouzn/EHD_Drop_3D}}}.

Moreover, our simulations confirmed that the dynamics of low-viscosity drops are strongly influenced by interfacial charge convection.
In this regime, the interplay between the flow and charge dynamics results in steep gradients in the interfacial charge density.
The development of these sharp features in the charge density profile results in the formation of ringing artifacts due to the Gibbs phenomenon.
Increasing the mesh resolution would eliminate the Gibbs phenomenon under relatively small electric capillary numbers $\mathrm{Ca_E}>0.2$.
However, the effect of nonlinear steepening by the flow is sufficiently strong at larger $\mathrm{Ca_E}$ that increasing the resolution does not avoid ringing.
To solve this problem, we introduced a weighted spherical harmonic transformation that serves as a relaxation method to exponentially damp high-wave-number coefficients.
The convergence of our numerical solution with decreasing level of relaxation shows that it provides a close approximation to the true solution in this regime.
Further characterizing the system's dynamics in this regime will require more sophisticated numerical methods for solutions containing discontinuities.

\section*{Acknowledgements}

The authors gratefully acknowledge funding from the US National Science Foundation under Grant No.\ CBET-1705377.

\addcontentsline{toc}{section}{\refname}\bibliography{Refs_shortjrtitle}

\providecommand{\noopsort}[1]{}\providecommand{\singleletter}[1]{#1}%
\begin{thebibliography}{63}%
\makeatletter
\providecommand \@ifxundefined [1]{%
 \@ifx{#1\undefined}
}%
\providecommand \@ifnum [1]{%
 \ifnum #1\expandafter \@firstoftwo
 \else \expandafter \@secondoftwo
 \fi
}%
\providecommand \@ifx [1]{%
 \ifx #1\expandafter \@firstoftwo
 \else \expandafter \@secondoftwo
 \fi
}%
\providecommand \natexlab [1]{#1}%
\providecommand \enquote  [1]{``#1''}%
\providecommand \bibnamefont  [1]{#1}%
\providecommand \bibfnamefont [1]{#1}%
\providecommand \citenamefont [1]{#1}%
\providecommand \href@noop [0]{\@secondoftwo}%
\providecommand \href [0]{\begingroup \@sanitize@url \@href}%
\providecommand \@href[1]{\@@startlink{#1}\@@href}%
\providecommand \@@href[1]{\endgroup#1\@@endlink}%
\providecommand \@sanitize@url [0]{\catcode `\\12\catcode `\$12\catcode
  `\&12\catcode `\#12\catcode `\^12\catcode `\_12\catcode `\%12\relax}%
\providecommand \@@startlink[1]{}%
\providecommand \@@endlink[0]{}%
\providecommand \url  [0]{\begingroup\@sanitize@url \@url }%
\providecommand \@url [1]{\endgroup\@href {#1}{\urlprefix }}%
\providecommand \urlprefix  [0]{URL }%
\providecommand \Eprint [0]{\href }%
\providecommand \doibase [0]{https://doi.org/}%
\providecommand \selectlanguage [0]{\@gobble}%
\providecommand \bibinfo  [0]{\@secondoftwo}%
\providecommand \bibfield  [0]{\@secondoftwo}%
\providecommand \translation [1]{[#1]}%
\providecommand \BibitemOpen [0]{}%
\providecommand \bibitemStop [0]{}%
\providecommand \bibitemNoStop [0]{.\EOS\space}%
\providecommand \EOS [0]{\spacefactor3000\relax}%
\providecommand \BibitemShut  [1]{\csname bibitem#1\endcsname}%
\let\auto@bib@innerbib\@empty
\bibitem [{\citenamefont {Basaran}\ \emph {et~al.}(2013)\citenamefont
  {Basaran}, \citenamefont {Gao},\ and\ \citenamefont
  {Bhat}}]{Basaran2013Ann_rev_inkjet}%
  \BibitemOpen
  \bibfield  {author} {\bibinfo {author} {\bibfnamefont {O.~A.}\ \bibnamefont
  {Basaran}}, \bibinfo {author} {\bibfnamefont {H.}~\bibnamefont {Gao}},\ and\
  \bibinfo {author} {\bibfnamefont {P.~P.}\ \bibnamefont {Bhat}},\ }\bibfield
  {title} {\bibinfo {title} {Nonstandard inkjets},\ }\href@noop {} {\bibfield
  {journal} {\bibinfo  {journal} {Annu. Rev. Fluid Mech.}\ }\textbf {\bibinfo
  {volume} {45}},\ \bibinfo {pages} {85} (\bibinfo {year} {2013})}\BibitemShut
  {NoStop}%
\bibitem [{\citenamefont {Fern{\'a}ndez~de
  La~Mora}(2007)}]{fernandez2007Ann_rev_spray}%
  \BibitemOpen
  \bibfield  {author} {\bibinfo {author} {\bibfnamefont {J.}~\bibnamefont
  {Fern{\'a}ndez~de La~Mora}},\ }\bibfield  {title} {\bibinfo {title} {The
  fluid dynamics of {Taylor} cones},\ }\href@noop {} {\bibfield  {journal}
  {\bibinfo  {journal} {Annu. Rev. Fluid Mech.}\ }\textbf {\bibinfo {volume}
  {39}},\ \bibinfo {pages} {217} (\bibinfo {year} {2007})}\BibitemShut
  {NoStop}%
\bibitem [{\citenamefont {Laser}\ and\ \citenamefont
  {Santiago}(2004)}]{laser2004review_micropumps}%
  \BibitemOpen
  \bibfield  {author} {\bibinfo {author} {\bibfnamefont {D.~J.}\ \bibnamefont
  {Laser}}\ and\ \bibinfo {author} {\bibfnamefont {J.~G.}\ \bibnamefont
  {Santiago}},\ }\bibfield  {title} {\bibinfo {title} {A review of
  micropumps},\ }\href@noop {} {\bibfield  {journal} {\bibinfo  {journal} {J.
  Micromech. Microeng.}\ }\textbf {\bibinfo {volume} {14}},\ \bibinfo {pages}
  {R35} (\bibinfo {year} {2004})}\BibitemShut {NoStop}%
\bibitem [{\citenamefont {Saville}(1997)}]{saville1997LDM_Ann_Rev}%
  \BibitemOpen
  \bibfield  {author} {\bibinfo {author} {\bibfnamefont {D.}~\bibnamefont
  {Saville}},\ }\bibfield  {title} {\bibinfo {title} {Electrohydrodynamics: the
  {Taylor-Melcher} leaky dielectric model},\ }\href@noop {} {\bibfield
  {journal} {\bibinfo  {journal} {Annu. Rev. Fluid Mech.}\ }\textbf {\bibinfo
  {volume} {29}},\ \bibinfo {pages} {27} (\bibinfo {year} {1997})}\BibitemShut
  {NoStop}%
\bibitem [{\citenamefont {Melcher}(1969)}]{melcher&taylor1969LDM}%
  \BibitemOpen
  \bibfield  {author} {\bibinfo {author} {\bibfnamefont {J.~R.}\ \bibnamefont
  {Melcher}},\ }\bibfield  {title} {\bibinfo {title} {Electrohydrodynamics: a
  review of the role of interfacial shear stresses},\ }\href@noop {} {\bibfield
   {journal} {\bibinfo  {journal} {Annu. Rev. Fluid Mech.}\ }\textbf {\bibinfo
  {volume} {1}},\ \bibinfo {pages} {111} (\bibinfo {year} {1969})}\BibitemShut
  {NoStop}%
\bibitem [{\citenamefont {Schnitzer}\ and\ \citenamefont
  {Yariv}(2015)}]{schnitzer2015taylor}%
  \BibitemOpen
  \bibfield  {author} {\bibinfo {author} {\bibfnamefont {O.}~\bibnamefont
  {Schnitzer}}\ and\ \bibinfo {author} {\bibfnamefont {E.}~\bibnamefont
  {Yariv}},\ }\bibfield  {title} {\bibinfo {title} {The {Taylor--Melcher} leaky
  dielectric model as a macroscale electrokinetic description},\ }\href@noop {}
  {\bibfield  {journal} {\bibinfo  {journal} {J. Fluid Mech.}\ }\textbf
  {\bibinfo {volume} {773}},\ \bibinfo {pages} {1} (\bibinfo {year}
  {2015})}\BibitemShut {NoStop}%
\bibitem [{\citenamefont {Mori}\ and\ \citenamefont
  {Young}(2018)}]{mori2018electrodiffusion}%
  \BibitemOpen
  \bibfield  {author} {\bibinfo {author} {\bibfnamefont {Y.}~\bibnamefont
  {Mori}}\ and\ \bibinfo {author} {\bibfnamefont {Y.~N.}\ \bibnamefont
  {Young}},\ }\bibfield  {title} {\bibinfo {title} {From electrodiffusion
  theory to the electrohydrodynamics of leaky dielectrics through the weak
  electrolyte limit},\ }\href@noop {} {\bibfield  {journal} {\bibinfo
  {journal} {J. Fluid Mech.}\ }\textbf {\bibinfo {volume} {855}},\ \bibinfo
  {pages} {67} (\bibinfo {year} {2018})}\BibitemShut {NoStop}%
\bibitem [{\citenamefont {Taylor}(1966)}]{taylor1966LDM}%
  \BibitemOpen
  \bibfield  {author} {\bibinfo {author} {\bibfnamefont {G.~I.}\ \bibnamefont
  {Taylor}},\ }\bibfield  {title} {\bibinfo {title} {Studies in
  electrohydrodynamics. {I. The} circulation produced in a drop by an electric
  field},\ }\href@noop {} {\bibfield  {journal} {\bibinfo  {journal} {Proc. R.
  Soc. A: Math. Phys. Eng. Sci.}\ }\textbf {\bibinfo {volume} {291}},\ \bibinfo
  {pages} {159} (\bibinfo {year} {1966})}\BibitemShut {NoStop}%
\bibitem [{\citenamefont {Ajayi}(1978)}]{ajayi1978note}%
  \BibitemOpen
  \bibfield  {author} {\bibinfo {author} {\bibfnamefont {O.~O.}\ \bibnamefont
  {Ajayi}},\ }\bibfield  {title} {\bibinfo {title} {A note on {Taylor’s}
  electrohydrodynamic theory},\ }\href@noop {} {\bibfield  {journal} {\bibinfo
  {journal} {Proc. R. Soc. A: Math. Phys. Eng. Sci.}\ }\textbf {\bibinfo
  {volume} {364}},\ \bibinfo {pages} {499} (\bibinfo {year}
  {1978})}\BibitemShut {NoStop}%
\bibitem [{\citenamefont {Zabarankin}(2013)}]{zabarankin2013liquid}%
  \BibitemOpen
  \bibfield  {author} {\bibinfo {author} {\bibfnamefont {M.}~\bibnamefont
  {Zabarankin}},\ }\bibfield  {title} {\bibinfo {title} {A liquid spheroidal
  drop in a viscous incompressible fluid under a steady electric field},\
  }\href@noop {} {\bibfield  {journal} {\bibinfo  {journal} {SIAM J. Appl.
  Math.}\ }\textbf {\bibinfo {volume} {73}},\ \bibinfo {pages} {677} (\bibinfo
  {year} {2013})}\BibitemShut {NoStop}%
\bibitem [{\citenamefont {Zhang}\ \emph {et~al.}(2013)\citenamefont {Zhang},
  \citenamefont {Zahn},\ and\ \citenamefont {Lin}}]{zhang2013transient}%
  \BibitemOpen
  \bibfield  {author} {\bibinfo {author} {\bibfnamefont {J.}~\bibnamefont
  {Zhang}}, \bibinfo {author} {\bibfnamefont {J.~D.}\ \bibnamefont {Zahn}},\
  and\ \bibinfo {author} {\bibfnamefont {H.}~\bibnamefont {Lin}},\ }\bibfield
  {title} {\bibinfo {title} {Transient solution for droplet deformation under
  electric fields},\ }\href@noop {} {\bibfield  {journal} {\bibinfo  {journal}
  {Phys. Rev. E}\ }\textbf {\bibinfo {volume} {87}},\ \bibinfo {pages} {043008}
  (\bibinfo {year} {2013})}\BibitemShut {NoStop}%
\bibitem [{\citenamefont {Lanauze}\ \emph {et~al.}(2013)\citenamefont
  {Lanauze}, \citenamefont {Walker},\ and\ \citenamefont
  {Khair}}]{lanauze2013inertia_EHD_drop}%
  \BibitemOpen
  \bibfield  {author} {\bibinfo {author} {\bibfnamefont {J.~A.}\ \bibnamefont
  {Lanauze}}, \bibinfo {author} {\bibfnamefont {L.~M.}\ \bibnamefont
  {Walker}},\ and\ \bibinfo {author} {\bibfnamefont {A.~S.}\ \bibnamefont
  {Khair}},\ }\bibfield  {title} {\bibinfo {title} {The influence of inertia
  and charge relaxation on electrohydrodynamic drop deformation},\ }\href@noop
  {} {\bibfield  {journal} {\bibinfo  {journal} {Phys. Fluids}\ }\textbf
  {\bibinfo {volume} {25}},\ \bibinfo {pages} {112101} (\bibinfo {year}
  {2013})}\BibitemShut {NoStop}%
\bibitem [{\citenamefont {Shkadov}\ and\ \citenamefont
  {Shutov}(2002)}]{shkadov2002drop}%
  \BibitemOpen
  \bibfield  {author} {\bibinfo {author} {\bibfnamefont {V.~Y.}\ \bibnamefont
  {Shkadov}}\ and\ \bibinfo {author} {\bibfnamefont {A.~A.}\ \bibnamefont
  {Shutov}},\ }\bibfield  {title} {\bibinfo {title} {Drop and bubble
  deformation in an electric field},\ }\href@noop {} {\bibfield  {journal}
  {\bibinfo  {journal} {Fluid Dyn.}\ }\textbf {\bibinfo {volume} {37}},\
  \bibinfo {pages} {713} (\bibinfo {year} {2002})}\BibitemShut {NoStop}%
\bibitem [{\citenamefont {Feng}(2002)}]{feng2002_2d_EHD}%
  \BibitemOpen
  \bibfield  {author} {\bibinfo {author} {\bibfnamefont {J.~Q.}\ \bibnamefont
  {Feng}},\ }\bibfield  {title} {\bibinfo {title} {A {2D} electrohydrodynamic
  model for electrorotation of fluid drops},\ }\href@noop {} {\bibfield
  {journal} {\bibinfo  {journal} {J. Colloid Interface Sci.}\ }\textbf
  {\bibinfo {volume} {246}},\ \bibinfo {pages} {112} (\bibinfo {year}
  {2002})}\BibitemShut {NoStop}%
\bibitem [{\citenamefont {He}\ \emph {et~al.}(2013)\citenamefont {He},
  \citenamefont {Salipante},\ and\ \citenamefont
  {Vlahovska}}]{he_salipate2013electrorotation}%
  \BibitemOpen
  \bibfield  {author} {\bibinfo {author} {\bibfnamefont {H.}~\bibnamefont
  {He}}, \bibinfo {author} {\bibfnamefont {P.~F.}\ \bibnamefont {Salipante}},\
  and\ \bibinfo {author} {\bibfnamefont {P.~M.}\ \bibnamefont {Vlahovska}},\
  }\bibfield  {title} {\bibinfo {title} {Electrorotation of a viscous droplet
  in a uniform direct current electric field},\ }\href@noop {} {\bibfield
  {journal} {\bibinfo  {journal} {Phys. Fluids}\ }\textbf {\bibinfo {volume}
  {25}},\ \bibinfo {pages} {032106} (\bibinfo {year} {2013})}\BibitemShut
  {NoStop}%
\bibitem [{\citenamefont {Das}\ and\ \citenamefont
  {Saintillan}(2017{\natexlab{a}})}]{das2017nonlinear}%
  \BibitemOpen
  \bibfield  {author} {\bibinfo {author} {\bibfnamefont {D.}~\bibnamefont
  {Das}}\ and\ \bibinfo {author} {\bibfnamefont {D.}~\bibnamefont
  {Saintillan}},\ }\bibfield  {title} {\bibinfo {title} {A nonlinear
  small-deformation theory for transient droplet electrohydrodynamics},\
  }\href@noop {} {\bibfield  {journal} {\bibinfo  {journal} {J. Fluid Mech.}\
  }\textbf {\bibinfo {volume} {810}},\ \bibinfo {pages} {225} (\bibinfo {year}
  {2017}{\natexlab{a}})}\BibitemShut {NoStop}%
\bibitem [{\citenamefont {Sherwood}(1988)}]{sherwood1988breakup}%
  \BibitemOpen
  \bibfield  {author} {\bibinfo {author} {\bibfnamefont {J.~D.}\ \bibnamefont
  {Sherwood}},\ }\bibfield  {title} {\bibinfo {title} {Breakup of fluid
  droplets in electric and magnetic fields},\ }\href@noop {} {\bibfield
  {journal} {\bibinfo  {journal} {J. Fluid Mech.}\ }\textbf {\bibinfo {volume}
  {188}},\ \bibinfo {pages} {133} (\bibinfo {year} {1988})}\BibitemShut
  {NoStop}%
\bibitem [{\citenamefont {Baygents}\ \emph {et~al.}(1998)\citenamefont
  {Baygents}, \citenamefont {Rivette},\ and\ \citenamefont
  {Stone}}]{baygents1998EHD}%
  \BibitemOpen
  \bibfield  {author} {\bibinfo {author} {\bibfnamefont {J.~C.}\ \bibnamefont
  {Baygents}}, \bibinfo {author} {\bibfnamefont {N.~J.}\ \bibnamefont
  {Rivette}},\ and\ \bibinfo {author} {\bibfnamefont {H.~A.}\ \bibnamefont
  {Stone}},\ }\bibfield  {title} {\bibinfo {title} {Electrohydrodynamic
  deformation and interaction of drop pairs},\ }\href@noop {} {\bibfield
  {journal} {\bibinfo  {journal} {J. Fluid Mech.}\ }\textbf {\bibinfo {volume}
  {368}},\ \bibinfo {pages} {359} (\bibinfo {year} {1998})}\BibitemShut
  {NoStop}%
\bibitem [{\citenamefont {Lac}\ and\ \citenamefont
  {Homsy}(2007)}]{lac2007axisymmetric}%
  \BibitemOpen
  \bibfield  {author} {\bibinfo {author} {\bibfnamefont {E.}~\bibnamefont
  {Lac}}\ and\ \bibinfo {author} {\bibfnamefont {G.~M.}\ \bibnamefont
  {Homsy}},\ }\bibfield  {title} {\bibinfo {title} {Axisymmetric deformation
  and stability of a viscous drop in a steady electric field},\ }\href@noop {}
  {\bibfield  {journal} {\bibinfo  {journal} {J. Fluid Mech.}\ }\textbf
  {\bibinfo {volume} {590}},\ \bibinfo {pages} {239} (\bibinfo {year}
  {2007})}\BibitemShut {NoStop}%
\bibitem [{\citenamefont {Lanauze}\ \emph {et~al.}(2015)\citenamefont
  {Lanauze}, \citenamefont {Walker},\ and\ \citenamefont
  {Khair}}]{lanauze2015EHD_JFM}%
  \BibitemOpen
  \bibfield  {author} {\bibinfo {author} {\bibfnamefont {J.~A.}\ \bibnamefont
  {Lanauze}}, \bibinfo {author} {\bibfnamefont {L.~M.}\ \bibnamefont
  {Walker}},\ and\ \bibinfo {author} {\bibfnamefont {A.~S.}\ \bibnamefont
  {Khair}},\ }\bibfield  {title} {\bibinfo {title} {Nonlinear
  electrohydrodynamics of slightly deformed oblate drops},\ }\href@noop {}
  {\bibfield  {journal} {\bibinfo  {journal} {J. Fluid Mech.}\ }\textbf
  {\bibinfo {volume} {774}},\ \bibinfo {pages} {245} (\bibinfo {year}
  {2015})}\BibitemShut {NoStop}%
\bibitem [{\citenamefont {Das}\ and\ \citenamefont
  {Saintillan}(2017{\natexlab{b}})}]{das2017EHD_simulation}%
  \BibitemOpen
  \bibfield  {author} {\bibinfo {author} {\bibfnamefont {D.}~\bibnamefont
  {Das}}\ and\ \bibinfo {author} {\bibfnamefont {D.}~\bibnamefont
  {Saintillan}},\ }\bibfield  {title} {\bibinfo {title} {Electrohydrodynamics
  of viscous drops in strong electric fields: numerical simulations},\
  }\href@noop {} {\bibfield  {journal} {\bibinfo  {journal} {J. Fluid Mech.}\
  }\textbf {\bibinfo {volume} {829}},\ \bibinfo {pages} {127} (\bibinfo {year}
  {2017}{\natexlab{b}})}\BibitemShut {NoStop}%
\bibitem [{\citenamefont {Hu}\ \emph {et~al.}(2015)\citenamefont {Hu},
  \citenamefont {Lai},\ and\ \citenamefont {Young}}]{Hu2015IBM_EHD_drop}%
  \BibitemOpen
  \bibfield  {author} {\bibinfo {author} {\bibfnamefont {W.~F.}\ \bibnamefont
  {Hu}}, \bibinfo {author} {\bibfnamefont {M.~C.}\ \bibnamefont {Lai}},\ and\
  \bibinfo {author} {\bibfnamefont {Y.~N.}\ \bibnamefont {Young}},\ }\bibfield
  {title} {\bibinfo {title} {A hybrid immersed boundary and immersed interface
  method for electrohydrodynamic simulations},\ }\href@noop {} {\bibfield
  {journal} {\bibinfo  {journal} {J. Comput. Phys.}\ }\textbf {\bibinfo
  {volume} {282}},\ \bibinfo {pages} {47} (\bibinfo {year} {2015})}\BibitemShut
  {NoStop}%
\bibitem [{\citenamefont {Bj{\o}rklund}(2009)}]{bjorklund2009level}%
  \BibitemOpen
  \bibfield  {author} {\bibinfo {author} {\bibfnamefont {E.}~\bibnamefont
  {Bj{\o}rklund}},\ }\bibfield  {title} {\bibinfo {title} {The level-set method
  applied to droplet dynamics in the presence of an electric field},\
  }\href@noop {} {\bibfield  {journal} {\bibinfo  {journal} {Comput. Fluids}\
  }\textbf {\bibinfo {volume} {38}},\ \bibinfo {pages} {358} (\bibinfo {year}
  {2009})}\BibitemShut {NoStop}%
\bibitem [{\citenamefont {Theillard}\ \emph {et~al.}(2019)\citenamefont
  {Theillard}, \citenamefont {Gibou},\ and\ \citenamefont
  {Saintillan}}]{theillard2019}%
  \BibitemOpen
  \bibfield  {author} {\bibinfo {author} {\bibfnamefont {M.}~\bibnamefont
  {Theillard}}, \bibinfo {author} {\bibfnamefont {F.}~\bibnamefont {Gibou}},\
  and\ \bibinfo {author} {\bibfnamefont {D.}~\bibnamefont {Saintillan}},\
  }\bibfield  {title} {\bibinfo {title} {Sharp numerical simulations of
  incompressible two-phase flows},\ }\href@noop {} {\bibfield  {journal}
  {\bibinfo  {journal} {J. Comput. Phys.}\ }\textbf {\bibinfo {volume} {391}},\
  \bibinfo {pages} {91} (\bibinfo {year} {2019})}\BibitemShut {NoStop}%
\bibitem [{\citenamefont {Feng}\ and\ \citenamefont
  {Scott}(1996)}]{feng1996FEM_EHD_drop}%
  \BibitemOpen
  \bibfield  {author} {\bibinfo {author} {\bibfnamefont {J.~Q.}\ \bibnamefont
  {Feng}}\ and\ \bibinfo {author} {\bibfnamefont {T.~C.}\ \bibnamefont
  {Scott}},\ }\bibfield  {title} {\bibinfo {title} {A computational analysis of
  electrohydrodynamics of a leaky dielectric drop in an electric field},\
  }\href@noop {} {\bibfield  {journal} {\bibinfo  {journal} {J. Fluid Mech.}\
  }\textbf {\bibinfo {volume} {311}},\ \bibinfo {pages} {289} (\bibinfo {year}
  {1996})}\BibitemShut {NoStop}%
\bibitem [{\citenamefont {Feng}(1999)}]{feng1999EHD_FEMdrop}%
  \BibitemOpen
  \bibfield  {author} {\bibinfo {author} {\bibfnamefont {J.~Q.}\ \bibnamefont
  {Feng}},\ }\bibfield  {title} {\bibinfo {title} {Electrohydrodynamic
  behaviour of a drop subjected to a steady uniform electric field at finite
  electric reynolds number},\ }\href@noop {} {\bibfield  {journal} {\bibinfo
  {journal} {Proc. R. Soc. A: Math. Phys. Eng. Sci.}\ }\textbf {\bibinfo
  {volume} {455}},\ \bibinfo {pages} {2245} (\bibinfo {year}
  {1999})}\BibitemShut {NoStop}%
\bibitem [{\citenamefont {Supeene}\ \emph {et~al.}(2008)\citenamefont
  {Supeene}, \citenamefont {Koch},\ and\ \citenamefont
  {Bhattacharjee}}]{supeene2008FEM_EHDdrop}%
  \BibitemOpen
  \bibfield  {author} {\bibinfo {author} {\bibfnamefont {G.}~\bibnamefont
  {Supeene}}, \bibinfo {author} {\bibfnamefont {C.~R.}\ \bibnamefont {Koch}},\
  and\ \bibinfo {author} {\bibfnamefont {S.}~\bibnamefont {Bhattacharjee}},\
  }\bibfield  {title} {\bibinfo {title} {Deformation of a droplet in an
  electric field: Nonlinear transient response in perfect and leaky dielectric
  media},\ }\href@noop {} {\bibfield  {journal} {\bibinfo  {journal} {J.
  Colloid Interface Sci.}\ }\textbf {\bibinfo {volume} {318}},\ \bibinfo
  {pages} {463} (\bibinfo {year} {2008})}\BibitemShut {NoStop}%
\bibitem [{\citenamefont {Collins}\ \emph {et~al.}(2013)\citenamefont
  {Collins}, \citenamefont {Sambath}, \citenamefont {Harris},\ and\
  \citenamefont {Basaran}}]{collins2013universal_scalinglaws}%
  \BibitemOpen
  \bibfield  {author} {\bibinfo {author} {\bibfnamefont {R.~T.}\ \bibnamefont
  {Collins}}, \bibinfo {author} {\bibfnamefont {K.}~\bibnamefont {Sambath}},
  \bibinfo {author} {\bibfnamefont {M.~T.}\ \bibnamefont {Harris}},\ and\
  \bibinfo {author} {\bibfnamefont {O.~A.}\ \bibnamefont {Basaran}},\
  }\bibfield  {title} {\bibinfo {title} {Universal scaling laws for the
  disintegration of electrified drops},\ }\href@noop {} {\bibfield  {journal}
  {\bibinfo  {journal} {PNAS}\ }\textbf {\bibinfo {volume} {110}},\ \bibinfo
  {pages} {4905} (\bibinfo {year} {2013})}\BibitemShut {NoStop}%
\bibitem [{\citenamefont {Wagoner}\ \emph {et~al.}(2021)\citenamefont
  {Wagoner}, \citenamefont {Vlahovska}, \citenamefont {Harris},\ and\
  \citenamefont {Basaran}}]{wagoner2021EHD_lenticular}%
  \BibitemOpen
  \bibfield  {author} {\bibinfo {author} {\bibfnamefont {B.~W.}\ \bibnamefont
  {Wagoner}}, \bibinfo {author} {\bibfnamefont {P.~M.}\ \bibnamefont
  {Vlahovska}}, \bibinfo {author} {\bibfnamefont {M.~T.}\ \bibnamefont
  {Harris}},\ and\ \bibinfo {author} {\bibfnamefont {O.~A.}\ \bibnamefont
  {Basaran}},\ }\bibfield  {title} {\bibinfo {title} {Electrohydrodynamics of
  lenticular drops and equatorial streaming},\ }\href@noop {} {\bibfield
  {journal} {\bibinfo  {journal} {J. Fluid Mech.}\ }\textbf {\bibinfo {volume}
  {925}} (\bibinfo {year} {2021})}\BibitemShut {NoStop}%
\bibitem [{\citenamefont {Veerapaneni}(2016)}]{veerapaneni2016BIM_EHD_vesicle}%
  \BibitemOpen
  \bibfield  {author} {\bibinfo {author} {\bibfnamefont {S.}~\bibnamefont
  {Veerapaneni}},\ }\bibfield  {title} {\bibinfo {title} {Integral equation
  methods for vesicle electrohydrodynamics in three dimensions},\ }\href@noop
  {} {\bibfield  {journal} {\bibinfo  {journal} {J. Comput. Phys.}\ }\textbf
  {\bibinfo {volume} {326}},\ \bibinfo {pages} {278} (\bibinfo {year}
  {2016})}\BibitemShut {NoStop}%
\bibitem [{\citenamefont {Sorgentone}\ \emph {et~al.}(2019)\citenamefont
  {Sorgentone}, \citenamefont {Tornberg},\ and\ \citenamefont
  {Vlahovska}}]{sorgentone20193d_EHD}%
  \BibitemOpen
  \bibfield  {author} {\bibinfo {author} {\bibfnamefont {C.}~\bibnamefont
  {Sorgentone}}, \bibinfo {author} {\bibfnamefont {A.~K.}\ \bibnamefont
  {Tornberg}},\ and\ \bibinfo {author} {\bibfnamefont {P.~M.}\ \bibnamefont
  {Vlahovska}},\ }\bibfield  {title} {\bibinfo {title} {A {3D} boundary
  integral method for the electrohydrodynamics of surfactant-covered drops},\
  }\href@noop {} {\bibfield  {journal} {\bibinfo  {journal} {J. Comput. Phys.}\
  }\textbf {\bibinfo {volume} {389}},\ \bibinfo {pages} {111} (\bibinfo {year}
  {2019})}\BibitemShut {NoStop}%
\bibitem [{\citenamefont {Sorgentone}\ \emph {et~al.}(2021)\citenamefont
  {Sorgentone}, \citenamefont {Kach}, \citenamefont {Khair}, \citenamefont
  {Walker},\ and\ \citenamefont {Vlahovska}}]{sorgentone2021drop_pair}%
  \BibitemOpen
  \bibfield  {author} {\bibinfo {author} {\bibfnamefont {C.}~\bibnamefont
  {Sorgentone}}, \bibinfo {author} {\bibfnamefont {J.~I.}\ \bibnamefont
  {Kach}}, \bibinfo {author} {\bibfnamefont {A.~S.}\ \bibnamefont {Khair}},
  \bibinfo {author} {\bibfnamefont {L.~M.}\ \bibnamefont {Walker}},\ and\
  \bibinfo {author} {\bibfnamefont {P.~M.}\ \bibnamefont {Vlahovska}},\
  }\bibfield  {title} {\bibinfo {title} {Numerical and asymptotic analysis of
  the three-dimensional electrohydrodynamic interactions of drop pairs},\
  }\href@noop {} {\bibfield  {journal} {\bibinfo  {journal} {J. Fluid Mech.}\
  }\textbf {\bibinfo {volume} {914}} (\bibinfo {year} {2021})}\BibitemShut
  {NoStop}%
\bibitem [{\citenamefont {Sorgentone}\ and\ \citenamefont
  {Vlahovska}(2022)}]{sorgentone2022tandem}%
  \BibitemOpen
  \bibfield  {author} {\bibinfo {author} {\bibfnamefont {C.}~\bibnamefont
  {Sorgentone}}\ and\ \bibinfo {author} {\bibfnamefont {P.~M.}\ \bibnamefont
  {Vlahovska}},\ }\bibfield  {title} {\bibinfo {title} {Tandem droplet
  locomotion in a uniform electric field},\ }\href@noop {} {\bibfield
  {journal} {\bibinfo  {journal} {arXiv e-prints}\ ,\ \bibinfo {pages} {arXiv}}
  (\bibinfo {year} {2022})}\BibitemShut {NoStop}%
\bibitem [{\citenamefont {Rallison}\ and\ \citenamefont
  {Acrivos}(1978)}]{rallison1978numerical}%
  \BibitemOpen
  \bibfield  {author} {\bibinfo {author} {\bibfnamefont {J.~M.}\ \bibnamefont
  {Rallison}}\ and\ \bibinfo {author} {\bibfnamefont {A.}~\bibnamefont
  {Acrivos}},\ }\bibfield  {title} {\bibinfo {title} {A numerical study of the
  deformation and burst of a viscous drop in an extensional flow},\ }\href@noop
  {} {\bibfield  {journal} {\bibinfo  {journal} {J. Fluid Mech.}\ }\textbf
  {\bibinfo {volume} {89}},\ \bibinfo {pages} {191} (\bibinfo {year}
  {1978})}\BibitemShut {NoStop}%
\bibitem [{\citenamefont {Pozrikidis}\ \emph {et~al.}(1992)\citenamefont
  {Pozrikidis} \emph {et~al.}}]{pozrikidis1992BIM_book}%
  \BibitemOpen
  \bibfield  {author} {\bibinfo {author} {\bibfnamefont {C.}~\bibnamefont
  {Pozrikidis}} \emph {et~al.},\ }\href@noop {} {\emph {\bibinfo {title}
  {Boundary integral and singularity methods for linearized viscous flow}}}\
  (\bibinfo  {publisher} {Cambridge university press},\ \bibinfo {year}
  {1992})\BibitemShut {NoStop}%
\bibitem [{\citenamefont {Zhao}\ \emph {et~al.}(2010)\citenamefont {Zhao},
  \citenamefont {Isfahani}, \citenamefont {Olson},\ and\ \citenamefont
  {Freund}}]{zhao_freund2010JCP}%
  \BibitemOpen
  \bibfield  {author} {\bibinfo {author} {\bibfnamefont {H.}~\bibnamefont
  {Zhao}}, \bibinfo {author} {\bibfnamefont {A.~H.~G.}\ \bibnamefont
  {Isfahani}}, \bibinfo {author} {\bibfnamefont {L.~N.}\ \bibnamefont
  {Olson}},\ and\ \bibinfo {author} {\bibfnamefont {J.~B.}\ \bibnamefont
  {Freund}},\ }\bibfield  {title} {\bibinfo {title} {A spectral boundary
  integral method for flowing blood cells},\ }\href@noop {} {\bibfield
  {journal} {\bibinfo  {journal} {J. Comput. Phys.}\ }\textbf {\bibinfo
  {volume} {229}},\ \bibinfo {pages} {3726} (\bibinfo {year}
  {2010})}\BibitemShut {NoStop}%
\bibitem [{\citenamefont {Bryngelson}\ and\ \citenamefont
  {Freund}(2018{\natexlab{a}})}]{bryngelson18b}%
  \BibitemOpen
  \bibfield  {author} {\bibinfo {author} {\bibfnamefont {S.~H.}\ \bibnamefont
  {Bryngelson}}\ and\ \bibinfo {author} {\bibfnamefont {J.~B.}\ \bibnamefont
  {Freund}},\ }\bibfield  {title} {\bibinfo {title} {Floquet stability analysis
  of capsules in viscous shear flow},\ }\href@noop {} {\bibfield  {journal}
  {\bibinfo  {journal} {J. Fluid Mech.}\ }\textbf {\bibinfo {volume} {852}},\
  \bibinfo {pages} {663} (\bibinfo {year} {2018}{\natexlab{a}})}\BibitemShut
  {NoStop}%
\bibitem [{\citenamefont {Bryngelson}\ and\ \citenamefont
  {Freund}(2019)}]{bryngelson19a}%
  \BibitemOpen
  \bibfield  {author} {\bibinfo {author} {\bibfnamefont {S.~H.}\ \bibnamefont
  {Bryngelson}}\ and\ \bibinfo {author} {\bibfnamefont {J.~B.}\ \bibnamefont
  {Freund}},\ }\bibfield  {title} {\bibinfo {title} {Non-modal {F}loquet
  stability of a capsule in large amplitude oscillatory extension},\
  }\href@noop {} {\bibfield  {journal} {\bibinfo  {journal} {Eur. J. Mech. B
  Fluids.}\ }\textbf {\bibinfo {volume} {77}},\ \bibinfo {pages} {171}
  (\bibinfo {year} {2019})}\BibitemShut {NoStop}%
\bibitem [{\citenamefont {Bryngelson}\ \emph {et~al.}(2019)\citenamefont
  {Bryngelson}, \citenamefont {Gu\'{e}niat},\ and\ \citenamefont
  {Freund}}]{bryngelson19b}%
  \BibitemOpen
  \bibfield  {author} {\bibinfo {author} {\bibfnamefont {S.~H.}\ \bibnamefont
  {Bryngelson}}, \bibinfo {author} {\bibfnamefont {F.}~\bibnamefont
  {Gu\'{e}niat}},\ and\ \bibinfo {author} {\bibfnamefont {J.~B.}\ \bibnamefont
  {Freund}},\ }\bibfield  {title} {\bibinfo {title} {Irregular dynamics of
  cellular blood flow in a model microvessel},\ }\href@noop {} {\bibfield
  {journal} {\bibinfo  {journal} {Phys. Rev. E}\ }\textbf {\bibinfo {volume}
  {100}},\ \bibinfo {pages} {012203} (\bibinfo {year} {2019})}\BibitemShut
  {NoStop}%
\bibitem [{\citenamefont {Bryngelson}\ and\ \citenamefont
  {Freund}(2018{\natexlab{b}})}]{bryngelson18a}%
  \BibitemOpen
  \bibfield  {author} {\bibinfo {author} {\bibfnamefont {S.~H.}\ \bibnamefont
  {Bryngelson}}\ and\ \bibinfo {author} {\bibfnamefont {J.~B.}\ \bibnamefont
  {Freund}},\ }\bibfield  {title} {\bibinfo {title} {Global stability of
  flowing red blood cell trains},\ }\href@noop {} {\bibfield  {journal}
  {\bibinfo  {journal} {Phys. Rev. Fluids}\ }\textbf {\bibinfo {volume} {3}}
  (\bibinfo {year} {2018}{\natexlab{b}})}\BibitemShut {NoStop}%
\bibitem [{\citenamefont {Freund}(2013)}]{freund2013flow}%
  \BibitemOpen
  \bibfield  {author} {\bibinfo {author} {\bibfnamefont {J.~B.}\ \bibnamefont
  {Freund}},\ }\bibfield  {title} {\bibinfo {title} {The flow of red blood
  cells through a narrow spleen-like slit},\ }\href@noop {} {\bibfield
  {journal} {\bibinfo  {journal} {Phys. Fluids}\ }\textbf {\bibinfo {volume}
  {25}},\ \bibinfo {pages} {110807} (\bibinfo {year} {2013})}\BibitemShut
  {NoStop}%
\bibitem [{\citenamefont {Firouznia}\ and\ \citenamefont
  {Saintillan}(2021)}]{firouznia2021PRF_EHD_film}%
  \BibitemOpen
  \bibfield  {author} {\bibinfo {author} {\bibfnamefont {M.}~\bibnamefont
  {Firouznia}}\ and\ \bibinfo {author} {\bibfnamefont {D.}~\bibnamefont
  {Saintillan}},\ }\bibfield  {title} {\bibinfo {title} {Electrohydrodynamic
  instabilities in freely suspended viscous films under normal electric
  fields},\ }\href@noop {} {\bibfield  {journal} {\bibinfo  {journal} {Phys.
  Rev. Fluids}\ }\textbf {\bibinfo {volume} {6}},\ \bibinfo {pages} {103703}
  (\bibinfo {year} {2021})}\BibitemShut {NoStop}%
\bibitem [{\citenamefont {Firouznia}\ \emph {et~al.}(2022)\citenamefont
  {Firouznia}, \citenamefont {Miksis}, \citenamefont {Vlahovska},\ and\
  \citenamefont {Saintillan}}]{firouznia2022JFM}%
  \BibitemOpen
  \bibfield  {author} {\bibinfo {author} {\bibfnamefont {M.}~\bibnamefont
  {Firouznia}}, \bibinfo {author} {\bibfnamefont {M.~J.}\ \bibnamefont
  {Miksis}}, \bibinfo {author} {\bibfnamefont {P.~M.}\ \bibnamefont
  {Vlahovska}},\ and\ \bibinfo {author} {\bibfnamefont {D.}~\bibnamefont
  {Saintillan}},\ }\bibfield  {title} {\bibinfo {title} {Instability of a
  planar fluid interface under a tangential electric field in a stagnation
  point flow},\ }\href@noop {} {\bibfield  {journal} {\bibinfo  {journal} {J.
  Fluid Mech.}\ }\textbf {\bibinfo {volume} {931}} (\bibinfo {year}
  {2022})}\BibitemShut {NoStop}%
\bibitem [{\citenamefont {Saad}\ and\ \citenamefont
  {Schultz}(1986)}]{saad1986gmres}%
  \BibitemOpen
  \bibfield  {author} {\bibinfo {author} {\bibfnamefont {Y.}~\bibnamefont
  {Saad}}\ and\ \bibinfo {author} {\bibfnamefont {M.~H.}\ \bibnamefont
  {Schultz}},\ }\bibfield  {title} {\bibinfo {title} {{GMRES}: A generalized
  minimal residual algorithm for solving nonsymmetric linear systems},\
  }\href@noop {} {\bibfield  {journal} {\bibinfo  {journal} {SIAM J. Sci.
  Comput.}\ }\textbf {\bibinfo {volume} {7}},\ \bibinfo {pages} {856} (\bibinfo
  {year} {1986})}\BibitemShut {NoStop}%
\bibitem [{\citenamefont {Boyd}(2001)}]{boyd2001chebyshev}%
  \BibitemOpen
  \bibfield  {author} {\bibinfo {author} {\bibfnamefont {J.~P.}\ \bibnamefont
  {Boyd}},\ }\href@noop {} {\emph {\bibinfo {title} {Chebyshev and Fourier
  spectral methods}}}\ (\bibinfo  {publisher} {Courier Corporation},\ \bibinfo
  {year} {2001})\BibitemShut {NoStop}%
\bibitem [{\citenamefont {Adams}\ and\ \citenamefont
  {Swarztrauber}(1999)}]{adams1999spherepack}%
  \BibitemOpen
  \bibfield  {author} {\bibinfo {author} {\bibfnamefont {J.~C.}\ \bibnamefont
  {Adams}}\ and\ \bibinfo {author} {\bibfnamefont {P.}~\bibnamefont
  {Swarztrauber}},\ }\bibfield  {title} {\bibinfo {title} {Spherepack 3.0: A
  model development facility},\ }\href@noop {} {\bibfield  {journal} {\bibinfo
  {journal} {MWR}\ }\textbf {\bibinfo {volume} {127}},\ \bibinfo {pages} {1872}
  (\bibinfo {year} {1999})}\BibitemShut {NoStop}%
\bibitem [{\citenamefont {Swarztrauber}\ and\ \citenamefont
  {Spotz}(2000)}]{swarztrauber2000generalized}%
  \BibitemOpen
  \bibfield  {author} {\bibinfo {author} {\bibfnamefont {P.~N.}\ \bibnamefont
  {Swarztrauber}}\ and\ \bibinfo {author} {\bibfnamefont {W.~F.}\ \bibnamefont
  {Spotz}},\ }\bibfield  {title} {\bibinfo {title} {Generalized discrete
  spherical harmonic transforms},\ }\href@noop {} {\bibfield  {journal}
  {\bibinfo  {journal} {J. Comput. Phys.}\ }\textbf {\bibinfo {volume} {159}},\
  \bibinfo {pages} {213} (\bibinfo {year} {2000})}\BibitemShut {NoStop}%
\bibitem [{\citenamefont {Rahimian}\ \emph {et~al.}(2015)\citenamefont
  {Rahimian}, \citenamefont {Veerapaneni}, \citenamefont {Zorin},\ and\
  \citenamefont {Biros}}]{rahimian2015boundary}%
  \BibitemOpen
  \bibfield  {author} {\bibinfo {author} {\bibfnamefont {A.}~\bibnamefont
  {Rahimian}}, \bibinfo {author} {\bibfnamefont {S.~K.}\ \bibnamefont
  {Veerapaneni}}, \bibinfo {author} {\bibfnamefont {D.}~\bibnamefont {Zorin}},\
  and\ \bibinfo {author} {\bibfnamefont {G.}~\bibnamefont {Biros}},\ }\bibfield
   {title} {\bibinfo {title} {Boundary integral method for the flow of vesicles
  with viscosity contrast in three dimensions},\ }\href@noop {} {\bibfield
  {journal} {\bibinfo  {journal} {J. Comput. Phys.}\ }\textbf {\bibinfo
  {volume} {298}},\ \bibinfo {pages} {766} (\bibinfo {year}
  {2015})}\BibitemShut {NoStop}%
\bibitem [{\citenamefont {Canuto}\ \emph {et~al.}(2012)\citenamefont {Canuto},
  \citenamefont {Hussaini}, \citenamefont {Quarteroni}, \citenamefont
  {Thomas~Jr} \emph {et~al.}}]{canuto2012spectral}%
  \BibitemOpen
  \bibfield  {author} {\bibinfo {author} {\bibfnamefont {C.}~\bibnamefont
  {Canuto}}, \bibinfo {author} {\bibfnamefont {M.~Y.}\ \bibnamefont
  {Hussaini}}, \bibinfo {author} {\bibfnamefont {A.}~\bibnamefont
  {Quarteroni}}, \bibinfo {author} {\bibfnamefont {A.}~\bibnamefont
  {Thomas~Jr}}, \emph {et~al.},\ }\href@noop {} {\emph {\bibinfo {title}
  {Spectral methods in fluid dynamics}}}\ (\bibinfo  {publisher} {Springer
  Science \& Business Media},\ \bibinfo {year} {2012})\BibitemShut {NoStop}%
\bibitem [{\citenamefont {Bruno}\ and\ \citenamefont
  {Kunyansky}(2001)}]{bruno2001FPOU_integration}%
  \BibitemOpen
  \bibfield  {author} {\bibinfo {author} {\bibfnamefont {O.~P.}\ \bibnamefont
  {Bruno}}\ and\ \bibinfo {author} {\bibfnamefont {L.~A.}\ \bibnamefont
  {Kunyansky}},\ }\bibfield  {title} {\bibinfo {title} {A fast, high-order
  algorithm for the solution of surface scattering problems: basic
  implementation, tests, and applications},\ }\href@noop {} {\bibfield
  {journal} {\bibinfo  {journal} {J. Comput. Phys.}\ }\textbf {\bibinfo
  {volume} {169}},\ \bibinfo {pages} {80} (\bibinfo {year} {2001})}\BibitemShut
  {NoStop}%
\bibitem [{\citenamefont {Ying}\ \emph {et~al.}(2006)\citenamefont {Ying},
  \citenamefont {Biros},\ and\ \citenamefont {Zorin}}]{ying2006polarpatch}%
  \BibitemOpen
  \bibfield  {author} {\bibinfo {author} {\bibfnamefont {L.}~\bibnamefont
  {Ying}}, \bibinfo {author} {\bibfnamefont {G.}~\bibnamefont {Biros}},\ and\
  \bibinfo {author} {\bibfnamefont {D.}~\bibnamefont {Zorin}},\ }\bibfield
  {title} {\bibinfo {title} {A high-order {3D} boundary integral equation
  solver for elliptic {PDEs} in smooth domains},\ }\href@noop {} {\bibfield
  {journal} {\bibinfo  {journal} {J. Comput. Phys.}\ }\textbf {\bibinfo
  {volume} {219}},\ \bibinfo {pages} {247} (\bibinfo {year}
  {2006})}\BibitemShut {NoStop}%
\bibitem [{\citenamefont {Veerapaneni}\ \emph {et~al.}(2011)\citenamefont
  {Veerapaneni}, \citenamefont {Rahimian}, \citenamefont {Biros},\ and\
  \citenamefont {Zorin}}]{veerapaneni2011fast}%
  \BibitemOpen
  \bibfield  {author} {\bibinfo {author} {\bibfnamefont {S.~K.}\ \bibnamefont
  {Veerapaneni}}, \bibinfo {author} {\bibfnamefont {A.}~\bibnamefont
  {Rahimian}}, \bibinfo {author} {\bibfnamefont {G.}~\bibnamefont {Biros}},\
  and\ \bibinfo {author} {\bibfnamefont {D.}~\bibnamefont {Zorin}},\ }\bibfield
   {title} {\bibinfo {title} {A fast algorithm for simulating vesicle flows in
  three dimensions},\ }\href@noop {} {\bibfield  {journal} {\bibinfo  {journal}
  {J. Comput. Phys.}\ }\textbf {\bibinfo {volume} {230}},\ \bibinfo {pages}
  {5610} (\bibinfo {year} {2011})}\BibitemShut {NoStop}%
\bibitem [{\citenamefont {Sorgentone}\ and\ \citenamefont
  {Tornberg}(2018)}]{sorgentone2018surfactant_JCP}%
  \BibitemOpen
  \bibfield  {author} {\bibinfo {author} {\bibfnamefont {C.}~\bibnamefont
  {Sorgentone}}\ and\ \bibinfo {author} {\bibfnamefont {A.~K.}\ \bibnamefont
  {Tornberg}},\ }\bibfield  {title} {\bibinfo {title} {A highly accurate
  boundary integral equation method for surfactant-laden drops in {3D}},\
  }\href@noop {} {\bibfield  {journal} {\bibinfo  {journal} {J. Comput. Phys.}\
  }\textbf {\bibinfo {volume} {360}},\ \bibinfo {pages} {167} (\bibinfo {year}
  {2018})}\BibitemShut {NoStop}%
\bibitem [{\citenamefont {Salipante}\ and\ \citenamefont
  {Vlahovska}(2010)}]{salipante_valhovska2010EHD_drop_PRF}%
  \BibitemOpen
  \bibfield  {author} {\bibinfo {author} {\bibfnamefont {P.~F.}\ \bibnamefont
  {Salipante}}\ and\ \bibinfo {author} {\bibfnamefont {P.~M.}\ \bibnamefont
  {Vlahovska}},\ }\bibfield  {title} {\bibinfo {title} {Electrohydrodynamics of
  drops in strong uniform dc electric fields},\ }\href@noop {} {\bibfield
  {journal} {\bibinfo  {journal} {Phys. Fluids}\ }\textbf {\bibinfo {volume}
  {22}},\ \bibinfo {pages} {112110} (\bibinfo {year} {2010})}\BibitemShut
  {NoStop}%
\bibitem [{\citenamefont {Ha}\ and\ \citenamefont
  {Yang}(2000)}]{ha2000EHD_quincke}%
  \BibitemOpen
  \bibfield  {author} {\bibinfo {author} {\bibfnamefont {J.~W.}\ \bibnamefont
  {Ha}}\ and\ \bibinfo {author} {\bibfnamefont {S.~M.}\ \bibnamefont {Yang}},\
  }\bibfield  {title} {\bibinfo {title} {Electrohydrodynamics and
  electrorotation of a drop with fluid less conductive than that of the ambient
  fluid},\ }\href@noop {} {\bibfield  {journal} {\bibinfo  {journal} {Phys.
  Fluids}\ }\textbf {\bibinfo {volume} {12}},\ \bibinfo {pages} {764} (\bibinfo
  {year} {2000})}\BibitemShut {NoStop}%
\bibitem [{\citenamefont {Sato}\ \emph {et~al.}(2006)\citenamefont {Sato},
  \citenamefont {Kaji}, \citenamefont {Mochizuki},\ and\ \citenamefont
  {Mori}}]{sato2006EHD_quincke}%
  \BibitemOpen
  \bibfield  {author} {\bibinfo {author} {\bibfnamefont {H.}~\bibnamefont
  {Sato}}, \bibinfo {author} {\bibfnamefont {N.}~\bibnamefont {Kaji}}, \bibinfo
  {author} {\bibfnamefont {T.}~\bibnamefont {Mochizuki}},\ and\ \bibinfo
  {author} {\bibfnamefont {Y.~H.}\ \bibnamefont {Mori}},\ }\bibfield  {title}
  {\bibinfo {title} {Behavior of oblately deformed droplets in an immiscible
  dielectric liquid under a steady and uniform electric field},\ }\href@noop {}
  {\bibfield  {journal} {\bibinfo  {journal} {Phys. Fluids}\ }\textbf {\bibinfo
  {volume} {18}},\ \bibinfo {pages} {127101} (\bibinfo {year}
  {2006})}\BibitemShut {NoStop}%
\bibitem [{\citenamefont {Quincke}(1896)}]{quincke1896ueber}%
  \BibitemOpen
  \bibfield  {author} {\bibinfo {author} {\bibfnamefont {G.}~\bibnamefont
  {Quincke}},\ }\bibfield  {title} {\bibinfo {title} {Ueber rotationen im
  constanten electrischen felde},\ }\href@noop {} {\bibfield  {journal}
  {\bibinfo  {journal} {Annalen der Physik}\ }\textbf {\bibinfo {volume}
  {295}},\ \bibinfo {pages} {417} (\bibinfo {year} {1896})}\BibitemShut
  {NoStop}%
\bibitem [{\citenamefont {Jones}(1984)}]{jones1984quincke}%
  \BibitemOpen
  \bibfield  {author} {\bibinfo {author} {\bibfnamefont {T.~B.}\ \bibnamefont
  {Jones}},\ }\bibfield  {title} {\bibinfo {title} {Quincke rotation of
  spheres},\ }\href@noop {} {\bibfield  {journal} {\bibinfo  {journal} {IEEE
  Trans. Ind. Appl.}\ ,\ \bibinfo {pages} {845}} (\bibinfo {year}
  {1984})}\BibitemShut {NoStop}%
\bibitem [{\citenamefont {Brosseau}\ and\ \citenamefont
  {Vlahovska}(2017)}]{brosseau2017streaming}%
  \BibitemOpen
  \bibfield  {author} {\bibinfo {author} {\bibfnamefont {Q.}~\bibnamefont
  {Brosseau}}\ and\ \bibinfo {author} {\bibfnamefont {P.~M.}\ \bibnamefont
  {Vlahovska}},\ }\bibfield  {title} {\bibinfo {title} {Streaming from the
  equator of a drop in an external electric field},\ }\href@noop {} {\bibfield
  {journal} {\bibinfo  {journal} {Phys. Rev. Lett.}\ }\textbf {\bibinfo
  {volume} {119}},\ \bibinfo {pages} {034501} (\bibinfo {year}
  {2017})}\BibitemShut {NoStop}%
\bibitem [{\citenamefont {Dommersnes}\ \emph {et~al.}(2013)\citenamefont
  {Dommersnes}, \citenamefont {Rozynek}, \citenamefont {Mikkelsen},
  \citenamefont {Castberg}, \citenamefont {Kjerstad}, \citenamefont {Hersvik},\
  and\ \citenamefont {Otto~Fossum}}]{dommersnes2013colloids_drop}%
  \BibitemOpen
  \bibfield  {author} {\bibinfo {author} {\bibfnamefont {P.}~\bibnamefont
  {Dommersnes}}, \bibinfo {author} {\bibfnamefont {Z.}~\bibnamefont {Rozynek}},
  \bibinfo {author} {\bibfnamefont {A.}~\bibnamefont {Mikkelsen}}, \bibinfo
  {author} {\bibfnamefont {R.}~\bibnamefont {Castberg}}, \bibinfo {author}
  {\bibfnamefont {K.}~\bibnamefont {Kjerstad}}, \bibinfo {author}
  {\bibfnamefont {K.}~\bibnamefont {Hersvik}},\ and\ \bibinfo {author}
  {\bibfnamefont {J.}~\bibnamefont {Otto~Fossum}},\ }\bibfield  {title}
  {\bibinfo {title} {Active structuring of colloidal armour on liquid drops},\
  }\href@noop {} {\bibfield  {journal} {\bibinfo  {journal} {Nat. Commun.}\
  }\textbf {\bibinfo {volume} {4}},\ \bibinfo {pages} {1} (\bibinfo {year}
  {2013})}\BibitemShut {NoStop}%
\bibitem [{\citenamefont {Ouriemi}\ and\ \citenamefont
  {Vlahovska}(2014)}]{ouriemi2014equatorial_vortices}%
  \BibitemOpen
  \bibfield  {author} {\bibinfo {author} {\bibfnamefont {M.}~\bibnamefont
  {Ouriemi}}\ and\ \bibinfo {author} {\bibfnamefont {P.~M.}\ \bibnamefont
  {Vlahovska}},\ }\bibfield  {title} {\bibinfo {title} {Electrohydrodynamics of
  particle-covered drops},\ }\href@noop {} {\bibfield  {journal} {\bibinfo
  {journal} {J. Fluid Mech.}\ }\textbf {\bibinfo {volume} {751}},\ \bibinfo
  {pages} {106} (\bibinfo {year} {2014})}\BibitemShut {NoStop}%
\bibitem [{\citenamefont {Gelb}(1997)}]{gelb1997gibbs_resolution_sh}%
  \BibitemOpen
  \bibfield  {author} {\bibinfo {author} {\bibfnamefont {A.}~\bibnamefont
  {Gelb}},\ }\bibfield  {title} {\bibinfo {title} {The resolution of the
  {Gibbs} phenomenon for spherical harmonics},\ }\href@noop {} {\bibfield
  {journal} {\bibinfo  {journal} {Math. Comput.}\ }\textbf {\bibinfo {volume}
  {66}},\ \bibinfo {pages} {699} (\bibinfo {year} {1997})}\BibitemShut
  {NoStop}%
\bibitem [{\citenamefont {Chung}\ \emph {et~al.}(2005)\citenamefont {Chung},
  \citenamefont {Robbins}, \citenamefont {Dalton}, \citenamefont {Davidson},
  \citenamefont {Alexander},\ and\ \citenamefont
  {Evans}}]{chung2005heatkernel_smoothing}%
  \BibitemOpen
  \bibfield  {author} {\bibinfo {author} {\bibfnamefont {M.~K.}\ \bibnamefont
  {Chung}}, \bibinfo {author} {\bibfnamefont {S.~M.}\ \bibnamefont {Robbins}},
  \bibinfo {author} {\bibfnamefont {K.~M.}\ \bibnamefont {Dalton}}, \bibinfo
  {author} {\bibfnamefont {R.~J.}\ \bibnamefont {Davidson}}, \bibinfo {author}
  {\bibfnamefont {A.~L.}\ \bibnamefont {Alexander}},\ and\ \bibinfo {author}
  {\bibfnamefont {A.~C.}\ \bibnamefont {Evans}},\ }\bibfield  {title} {\bibinfo
  {title} {Cortical thickness analysis in autism with heat kernel smoothing},\
  }\href@noop {} {\bibfield  {journal} {\bibinfo  {journal} {NeuroImage}\
  }\textbf {\bibinfo {volume} {25}},\ \bibinfo {pages} {1256} (\bibinfo {year}
  {2005})}\BibitemShut {NoStop}%
\end{thebibliography}%


\end{document}